\documentclass[12pt,preprint]{aastex}

\newcommand{\ang}{\AA}
\newcommand{\teff}{\ensuremath{T_{\rm eff}}}
\newcommand{\logg}{\ensuremath{\log (g)}}
\newcommand{\kms}{\,km~s$^{-1}$}
\newcommand{\subsun}{\mbox{$_{\normalsize\odot}$}}

\shortauthors{Schweitzer et al.}

\begin{document}

\title{Analysis of Keck HIRES spectra of early L-type dwarfs}

\author{Andreas Schweitzer}
\affil{Department of Physics and Astronomy \& Center for Simulational Physics, University of Georgia, Athens, GA 30602-2451 \email{andy@physast.uga.edu}}
    
\author{John E. Gizis}
\affil{Infrared Processing and Analysis Center, 100-22,
  California Institute of Technology, 
  Pasadena, CA 91125\email{gizis@ipac.caltech.edu}}

\author{Peter H. Hauschildt}
\affil{Department of Physics and Astronomy \& Center for Simulational Physics, University of Georgia, Athens, GA 30602-2451\email{yeti@hobbes.physast.uga.edu}}

\author{France Allard}
\affil{CRAL, Ecole Normale Superieure, 46 Alle d'Italie, Lyon, 69364, France \email{fallard@ens-lyon.fr}}

\and

\author{I. Neill Reid}
\affil{Department of  Physics and Astronomy, University of Pennsylvania,
209 South 33rd Street, Philadelphia PA 19104-6396 \email{inr@hep.upenn.edu}}

\begin{abstract}
We present analyses of high resolution and medium resolution spectra of
early L dwarfs. We used our latest set of model atmospheres to reproduce
and analyze the observed features. We can model the optical flux and
the atomic line profiles with the best accuracy to date.
The models used to reproduce the observations include dust condensation and
dust opacities. Compared to previous studies using older models we
find that our dust treatment is much improved.
The derived parameters for the objects are well in the expected range
for old very low mass objects.
This is also supported by the absence of Li in most of the objects.
For the objects showing Li we can be almost certain that those
are brown dwarfs. However, a spectral analysis in general, and this
one in particular can only very roughly determine mass and age.
\end{abstract}


\section{Introduction}

In recent years many new very low mass objects have been discovered, classified and
the techniques to detect and observe them have been refined.
With the increased sample size has come the discovery of 
bright sources, whose characteristics can be assumed to be representative 
of the underlying population. Those bright sources are accessible to
study at high resolution and high signal-to-noise.
Such investigations can now enable us to measure their parameters which is
necessary if we want to understand the physics and nature of very low mass
objects in general.

The models for such objects have experienced an equally rapid
evolution in the last few years.
The latest model atmospheres incorporate the presence of dust within the
atmosphere.
Although it has become clear that dust does form in cool atmospheres
a realistic reproduction of the observed effects has not been achieved yet.
The problems associated with implementing an accurate treatment of dust
in cool dwarf atmospheres include devising consistent formation and  
destruction mechanisms, including condensation and diffusion, besides 
incorporating a realistic treatment of opacities.

Earlier work on the interpretation of L~dwarf spectra
\citep[e.g.][]{burrows2000,pavlenko2000,leinert2000}
focused either on the reproduction and explanation of individual
features or used medium to low resolution spectra.
\citet{basri2000} analyzed high resolution spectra of a sample of
L~dwarfs using a first generation
of the dusty and rained out models by \citet{ara97}.  The dusty
models used in this
earlier work seemed to overestimate the effects of the dust on the
optical line formation.
The rained out models appeared to provide an adequate fit to the lines.
However, it was 
clear from other work by \citet{ruiz97},   \citet{leggett98},
and \citet{kirk2000gd}
that these L type dwarfs share the near-infrared properties of dusty
models and not those
of rained out models.   

Recently, \citet{cond-dusty} computed an improved set of dusty and
rained out models 
which include revised opacity and dust treatment.
These models were used by \citet{lsandy}
to analyse low
and medium resolution spectra of the overall spectral distribution of M
and L dwarfs.  Their
results confirmed that the new models could still reproduce the
near-infrared properties of
L dwarfs within uncertainties of the water vapor opacity profile, but
also showed that they
fitted adequately their optical spectral distributions.

In this paper we will use the improved models of \citet{cond-dusty}
which were also used in \citet{lsandy} to
perform analyses of optical high resolution spectra similarly to
\citet{basri2000}.
In the next section we will describe the available data,
in Section \ref{modelsec} we will describe the models and the
important improvements,
Section \ref{anasec} contains the methods
we used to analyze the data and the results.

\section{Observations and data}

Echelle spectroscopy of the L dwarfs discussed in this paper
was obtained using the HIRES spectrograph \citep{hires} on the
Keck I 10-meter telescope. The observations were made on 23 and 24 August 
1998, 13 November 1998 and 6 and 7 June 1999. In each case we used the
red cross-disperser, covering selected regions of the spectrum in
the 6000 to 9800\AA\ wavelength region. The August observations were
made using the $1".15 \times 7".0$ slit, giving a resolution of
$\sim34,000$ (8.8 kms$^{-1}$ or 4 pixels); seeing was sufficiently better
than 1 arcsecond
on the other three nights to allow use of the $0".86 \times 7".0$
slit, giving an increased resolution of $\sim45,000$ (6.4 kms$^{-1}$ or
3 pixels). Typical exposure times were 3000 seconds.

The echelle data were extracted using the {\sl mckee} suite of programs, written by 
T. Barlow. Most of the L dwarfs are barely detected at wavelengths shortward of
$\sim7000$\AA\ (sometimes $\lambda < 8000$\AA), so we used observations of brighter,
near-featureless stars (usually white dwarfs) as templates to outline the shape of
each order. Similar techniques were used to extract the thorium-argon wavelength
calibration spectra, obtained at the start and end of each night, 
while flat-field images are used to correct gross non-uniformities. 
The data were wavelength calibrated using standard IRAF routines and the
thorium-argon spectra.
We have not attempted to set the spectra on an absolute flux scale. 

The low resolution spectra are data published in \citet{kirk99}
and \citet{kirk2000}.

\section{The models}
\label{modelsec}
The models used in this work are the ones described in \citet{cond-dusty}.
A detailed description can be found therein.
They are the same as in the earlier work of \citet{lsandy}.
The models utilize the
Ames H$_2$O  and TiO line lists by  \citet{ames-water-new} and \citet{ames-tio}.
We  stress  that large
uncertainties persist  in these line data for the  temperature range of
this work \citep[see][]{allard2000}.  The opacities in
the models  have been upgraded with
(i) the  replacement  of  the  JOLA  (Just Overlapping  Line
Approximation) opacities for FeH, VO  and CrH by theoretical line
data for which we do dynamical opacity sampling
as described in \citet{cond-dusty},
(ii)  the extension of our database  of dust grain
opacities from 10  to 40 species which are based on laboratory 
measurements and (iii) the revision
of pressure induced H$_2$ opacities.

The dust for all the species of \citet{sharp_huebner1990} forms in chemical equilibrium
using the Gibbs
free energy from the JANAF database \citep{janaf}.
The models treat the dust in the limiting case where the dust is in
equilibrium with the gas phase. When dust forms in a layer
it does not diffuse out of the layer but contributes in full to the opacity (Ames-Dusty-2000 models).
As comparison we also used the other extreme where the dust
forms and completely settles below the atmosphere (Ames-Cond models) and, therefore, does no longer
contribute to the opacity.
A detailed description of the treatment of dust and of the effects of dust
on the temperature structure and the spectrum can be found
in \citet{cond-dusty}.

The temperature structures of these models were used to calculate
synthetic high resolution spectra at a resolution of 0.1\ang\ between
6400 and 9100\ang\ for all the models used in this analysis.
The atomic lines in the spectra are dominated by van~der~Waals
pressure broadening which has been taken into account as in
the calculations as described in \citet{schweitzer96}.
Although not part of the targeted wavelength range, our
calculations naturally include all strong lines whose wings end up
in the targeted wavelength range \citep[see e.g.][]{NGhot}.
This includes most
prominently the wings of the very strong Na~D doublet.
For the fitting process, the spectra were rotationally broadened if
necessary.
They were then convolved with a Gaussian to match the resolution
of the observed HIRES spectra.

We include van~der~Waals (vdW) broadening since the lines are pressure
broadened due to collisions with neutral species.
We use a modification of the Uns{\"o}ld treatment of vdW broadening
which considers not only H but also H$_2$, He and the most abundant
trace species \citep[see][]{schweitzer96}.
In a different study, \citet{burrows2000} deploy an approximation for
pressure broadening due to H$_2$ which is an extension to the
impact approximation used by the Uns{\"o}ld approach.
For our model structures we calculated the Weisskopf radius which distinguishes
between the validity of the impact theory and the statistical theory of
vdW broadening \citep[see][]{burrows2000} but found that the
impact theory is adequate.
This was confirmed by other tests we have made using different line
broadening profiles like Margenau profiles \citep{margenau32}, which   
is a general approach to the statistical theory and proved worse (see
section \ref{highresana} for the high resolution fits).
However, the models presented by \citet{burrows2000}
(which do not use a plain Uns{\"o}ld treatment)
do not include the molecular line data present in our models and,
as a result, predict substantially different spectral energy
distributions. Both sets of models therefore have their limitations.
We aim to integrate more sophisticated broadening calculations,
especially for the treatment of the far wings of resonance lines,
in a future generation of stellar models.

However, the most important impact on line profiles is caused
by the presence of dust. This has already been demonstrated
and investigated by \citet{leinert2000} and \citet{basri2000}
where Dusty and Cond models
have been compared in reproducing individual lines. The line width
is very sensitive to the amount of dust as a function of layer.
Errors on line broadening theories become negligible compared
to the effect of dust and that justifies why the former should only be
addressed once the dust issue is settled.

\section{Analysis and results}
\label{anasec}

We used a $\chi^2$ fitting technique to determine the
best fitting model spectra. We used a grid of model spectra and
for each spectrum we calculated a $\chi^2$ in order
to determine the best fitting model which had the smallest
$\chi^2$.
All spectra were visually inspected. Some fits had to be corrected
when the best fitting model was a  feature poor spectrum
and the fit was simply an overall canceling out of the observed
features in the $\chi^2$ determination.
Excluding the problematic model from the grid of theoretical
models usually corrected this problem.
We did this for fitting both the low resolution models and the
high resolution models.
Our grid of theoretical spectra consisted of solar metallicity models
with effective temperatures between
1600~K and 2400~K in steps of 100~K and surface gravities with \logg\ between 3.5 and 5.5
in steps of 0.5.

\subsection{Results for the low resolution spectra}
We used the low resolution spectra to determine a first fit of the
effective temperature.
We compared the effect of \logg\ for constant effective temperature and found that
the gravity does not significantly change the shape of the optical
low-resolution spectra.
This is also supported by the results of our $\chi^2$ fitting 
techniques which finds similar effective temperatures rather than
similar gravities as the three best fitting models.
We restricted the wavelength range to 7700\ang\ to 9900\ang\ to
avoid the problematic VO band on the blue side of the wavelength range
and the problematic FeH band on the red side.
For both of these bands we use 
the latest not fully tested available data.

The results of the fitting process is summarized in Tab. \ref{lowresfittab}
The derived parameters have a determination uncertainty of 100~K in effective temperature and
0.5~dex in \logg.
These errors reflect the spacing of the grid.
The $\chi^2$ fitting
process also determines the second and third best fits within 100~K
and 0.5~dex.
These errors do not include any systematic problems in the models
and the values have to be considered as determined within
the used models.
The table lists the fits for both the AMES-Dusty and the AMES-Cond models.
Except for the L3.5 dwarf, the Dusty models give
better fits when inspected visually.
With the AMES-Cond models the \ion{K}{1} is too strong and
the CrH and FeH bands at 8611\ang\ and 8692\ang\ are too strong.
This is still the case for the L3.5 dwarf, however, for all parameters,
the \ion{K}{1} doublet is too weak. This gives a hint that at these
temperatures models between the limiting cases of
AMES-Dusty and AMES-Cond start being more appropriate.
Such models that calculate the amount of dust which
diffuses are being developed (Allard et al., in preparation).
They will be needed for effective temperatures below
approximately 1800~K.
In this paper the dwarfs are reasonably hot enough that
such models are not required.
The fact that we can use the limiting AMES-Dusty models is
also supported by the results of \citet{lsandy}; and
\citet{chabrier2000} suggests that such atmosphere conditions
exist.

The fits to the low resolution spectra usually return high gravity values.
This is because the high gravity spectra have very broad
atomic lines which smooth the whole spectrum.
These spectra are therefore preferred by the  $\chi^2$ test.

The fits of AMES-Dusty models to the low resolution
spectra are shown in Figs. \ref{lowfitplot0149} through \ref{lowfitplot0036}.
As can be seen the molecular bands are well reproduced.
This is true for VO, CrH and FeH  which produce the various features and
also for TiO which determines the overall
optical spectrum.
No attempt has been made to fit individual strong atomic lines
like the \ion{K}{1} resonance doublet or the \ion{Na}{1} subordinate
doublet. The purpose of these fits was to determine the
effective temperature by fitting the overall shape which
is dominated by molecular absorption.
The region around the \ion{Na}{1} doublet usually can be fit by scaling
the spectra but this would worsen the fit elsewhere.
Similarly for the {\ion{K}{1} doublet, which could
be fit better by slightly different parameters.
This problem will be addressed below when discussing
the differences between the results from high resolution
and low resolution spectra.

As expected  and within the error bars our fits confirm a correlation of spectral
type with effective temperature.
The overall temperature range of only
300~K for 4 spectral classes is not that small when translating the effective
temperature back into luminosity (which it originally defines).
However, the small temperature range is also not very surprising since
dust condensation (and the associated molecule depletion) takes
place very rapidly in parameter space. Therefore, molecular
features are very sensitive to changes in \teff.

\subsection{Results for the high resolution spectra}
\label{highresana}

We fitted model spectra to the \ion{Cs}{1} line at 8521.36~\ang, the \ion{Rb}{1} lines
at 7800.26~\ang\ and 7947.60~\ang, the \ion{Na}{1} doublet at 8190\ang\ and the \ion{K}{1}
at 7685\ang\ doublet when these lines were
in the observed wavelength range. For two spectra we also used the \ion{Ca}{1} line 
at 6572.77~\ang\ and
for one we used the \ion{Li}{1} line at 6707.76~\ang.
The results are tabulated in Tab. \ref{highresfittab}. The table also demonstrates which
lines could be observed in which object.
The fits to the lines are shown in Figs. \ref{hi0149fit}--\ref{hi0036fit}.

As can be seen the derived temperatures do not correlate with
spectral type and the temperature derived from the low resolution
spectra. The error bars are again 100~K and 0.5~dex within
the previously established limits.
The derived temperatures from low resolution and high resolution
spectra don not agree for the later dwarfs in the sample.
Only the earlier type objects show a weak correlation between effective temperatures
derived from low resolution  and high resolution spectra.
To prevent a discrepancy between low resolution and high resolution spectra
 we could have fixed the effective temperature to the one
derived from the low resolution spectra during the fitting process.
We attempted this but the resulting
fits were very poor so that we left the effective temperature as a
free parameter.

The reason for the discrepancy between effective temperatures
derived from low resolution spectra and high resolution
spectra is most likely a 
dust opacity and dust settling problem \citep[see also][]{leinert2000}.
A combination of temperature structure and opacity structure 
produces a line that ``looks'' too hot for the later dwarfs.
This is another indicator that the cooler the models get, the less satisfied
the conditions are that allow the use of the AMES-Dusty models.
Usage of more consistent models will first change only parts of the atmosphere
before the whole atmosphere and spectrum changes.
The molecular bands form in different regions and will be affected differently
by partial incorrectness than atomic lines which usually form in the deepest layers.

For each object, the values of \logg\ are usually within $\pm$0.5~dex.
However, all objects have rather large values for \logg\ although the
used parameter range started as low as \logg=3.5.
This means that all objects have already contracted considerably.
This is also supported by the fact that only two objects show the
\ion{Li}{1} line. We discuss this in detail in section \ref{licorr}.

All the models had been rotationally broadened if necessary.
We did not include the rotationally velocity in the automatic
fitting process. We rather adjusted this parameter by visual
inspection. We want to note, however, that changing the
rotational velocity by 5 or 10 \kms\ did usually change the
quality of the fit, but it did not change the parameters of the
best fitting models. Therefore, the accuracy of the rotational
broadening is not important for the determination of \teff\ and
\logg. Keeping these caveats in mind we give the best fitting values for the rotational velocity
in Tab. \ref{highresfittab}.

\subsection{2MASS0345+2540}
2MASS0345+2540 was also in the sample of \citet{lsandy} who found \teff=2000~K and \logg=6.0,
whereas we find \teff=1900~K and \logg=5.5. This is well consistent within the error bars.
It is however to note that \citet{lsandy} used the full spectral range from the optical to the
NIR.
The current data for water still has uncertainties \citep{allard2000} as it does not
reproduce the H-band correctly.
This affects the overall flux distribution and the optical will
have a different pseudo-continuum level.
An unrealistically shaped overall spectrum will also affect the
pure technical fitting process which will find the best compromise.
Both these effects will lead to systematic differences between
this work and the work of \citet{lsandy}.

This example shows that there is an unknown systematic error in addition to
the the error bars from our measurement.
Using a larger wavelength range and using better line opacities (which is
not possible at the moment) may change the results.

\subsection{Li presence and substellar nature}
\label{licorr}
We only derived high ($\gtrsim4.5$) values for \logg.
This is consistent with the fact that only the two objects
2MASS1146+2230 and 2MASS1726+1538 show a \ion{Li}{1} line.
The \ion{Li}{1} line in 2MASS1726+1538 is very weak \citep[see][]{kirk2000} and is
not detected in the HIRES observation.
To demonstrate this consistency we plot \teff\ versus \logg\ from the data by 
\citet{chabrier2000} in Fig. \ref{teff_logg}.
The dotted lines indicate 95\% Li abundance remaining (left
line) and 50\% Li abundance remaining (right line).
As can be seen, objects without Li should all have \logg\ above 5.0,
which is what we derived.
2MASS1146+2230
will then have to have a \logg\ in the lower
part of its error bars and is almost certainly a brown dwarf.
2MASS1726+1538 has a very weak \ion{Li}{1} line and is, therefore,
in the strip where Lithium depletion occurs.
In the context of these models, 2MASS1726+1538 then has \logg=5.25 and
m$\approx$0.06m\subsun\ and would be a brown dwarf as well.
All other objects are hot enough to be either brown dwarfs or
stars.
The tracks below \logg=4.5 correspond to masses below 0.03~M\subsun\ and
ages significantly younger than 100~Myrs. Such objects are unlikely to
be found in a field sample such as 2MASS.

We want to emphasize that the derived error bars are much too large
to use the spectroscopic parameters to determine mass and age from
such a diagram. We use this diagram for illustrative purposes and to
show that we are consistent with what is expected from evolutionary models.

\section{Discussion and summary}

We have presented fits to high and medium resolution spectra of L~dwarfs
using
our latest model atmospheres and
synthetic spectra.

The observed spectra including the high resolution line profiles
can be reproduced very well.
The derived parameters from the best fitting models are physically reasonable.
They match the expected parameters for old ($\gtrsim$ 1Gyr) VLM stars and massive brown dwarfs.

The important physics in the models is the inclusion of dust condensation
and dust opacity as described above.
The models used in this analysis use the limiting case in which the dust forms
and does not diffuse.
Since most objects can be well reproduced with these models,
this assumption seems to be well justified for early L~dwarfs.
However, we found hints that for objects later than L3 (or with
our models cooler than 1800~K) the assumption for the 
limiting case is no longer
fully met \citep[see also][]{lsandy}.

Most importantly, however, within the valid range our models reproduce
accurately the observed spectra.
Our new models reproduce realistic line profiles in the presence
of dust.
The applied Uns{\"o}ld treatment for the vdW broadening is only an
approximation. But for the objects in this paper it reproduces the atomic
lines very well within the limits of the dust distribution uncertainties.
Tests using other profiles did not reproduce better profiles.
We are actively improving the models and considering more
realistic line broadening mechanism similar to \citet{burrows2000} but
find no reason to change the current implementation yet.

We encountered the problem that the determined effective temperatures 
from high resolution spectra are different than those from low resolution spectra.
We think this could have two reasons.
The first one could be that the line formation of the atomic lines
is not reproduced correctly due to an incorrect temperature or opacity structure.
This can easily happen if the dust treatment does not produce the right dust
properties for all layers
\citep[see also][]{leinert2000}.
The second reason is simply the fact that the pseudo continuum level in the
optical is probably not correct due to incorrect or incomplete
line data.
If the molecular line data is incorrect then
the overall flux is not distributed correctly which will lead
to different pseudo continuum around the atomic lines and thus different
best fitting parameters for atomic lines.

Both of these effects can be interpreted as seeing hotter (or cooler)
temperatures in the atomic lines. Therefore, we think that the
distortions on the \logg\ determinations are not very large and that
our derived \logg\ values are not affected very much.
This is also supported by the fact that we 
find high values
for \logg\ for objects
without Li (see above).

Future models, which are being developed, will include a more
consistent dust treatment which will also be valid for lower 
temperatures and later L~dwarfs. The improved dust treatment
plus more complete and improved line data which becomes
available over time will make the models applicable to
early and late type L~dwarfs at the same time and it
will give consistent results for low resolution and
high resolution spectra.

\acknowledgments
We are
grateful to Richard Freedman (NASA-Ames) who generously provided VO and
CrH line lists for use in the current 
models.
AS acknowledges support from NASA ATP grant NAG 5-8425 to
the University of Georgia.
JEG acknowledges the
support of the Jet Propulsion Laboratory, California
Institute of Technology, which is operated under contract
with NASA.
PHH acknowledges support in  part from NASA
ATP grant  NAG 5-3018 and LTSA  grant NAG 5-3619 to  the University of 
Georgia and
partial support from the P\^ole Scientifique de
Mod\'elisation Num\'erique at ENS-Lyon.
FA  acknowledges support from CNRS.
INR acknowledges partial support from a NASA/JPL grant
to 2MASS Core Project Science.
This work was also supported
in part by NSF grants AST-9417242, AST-9731450, and NASA grant
NAG5-3505 and an IBM SUR grant to the University of Oklahoma. Some of 
the calculations presented in this paper were performed on the IBM SP
and the SGI Origin 2000 of the UGA UCNS, on the IBM SP of the San
Diego Supercomputer Center (SDSC, with support from the National
Science Foundation), on the Cray T3E and the IBM SP of the NERSC with support from
the DoE, on the IBM SP2 of the French Centre National Universitaire
Sud de Calcul (CNUSC).  We thank all these institutions for a generous
allocation of computer time.
This paper is partially based on observations obtained at the
W.~M. Keck Observatory, which is operated jointly by the
University of California, the California Institute of Technology and NASA.

\newpage

\begin{figure}
\epsscale{0.49}
\plotone{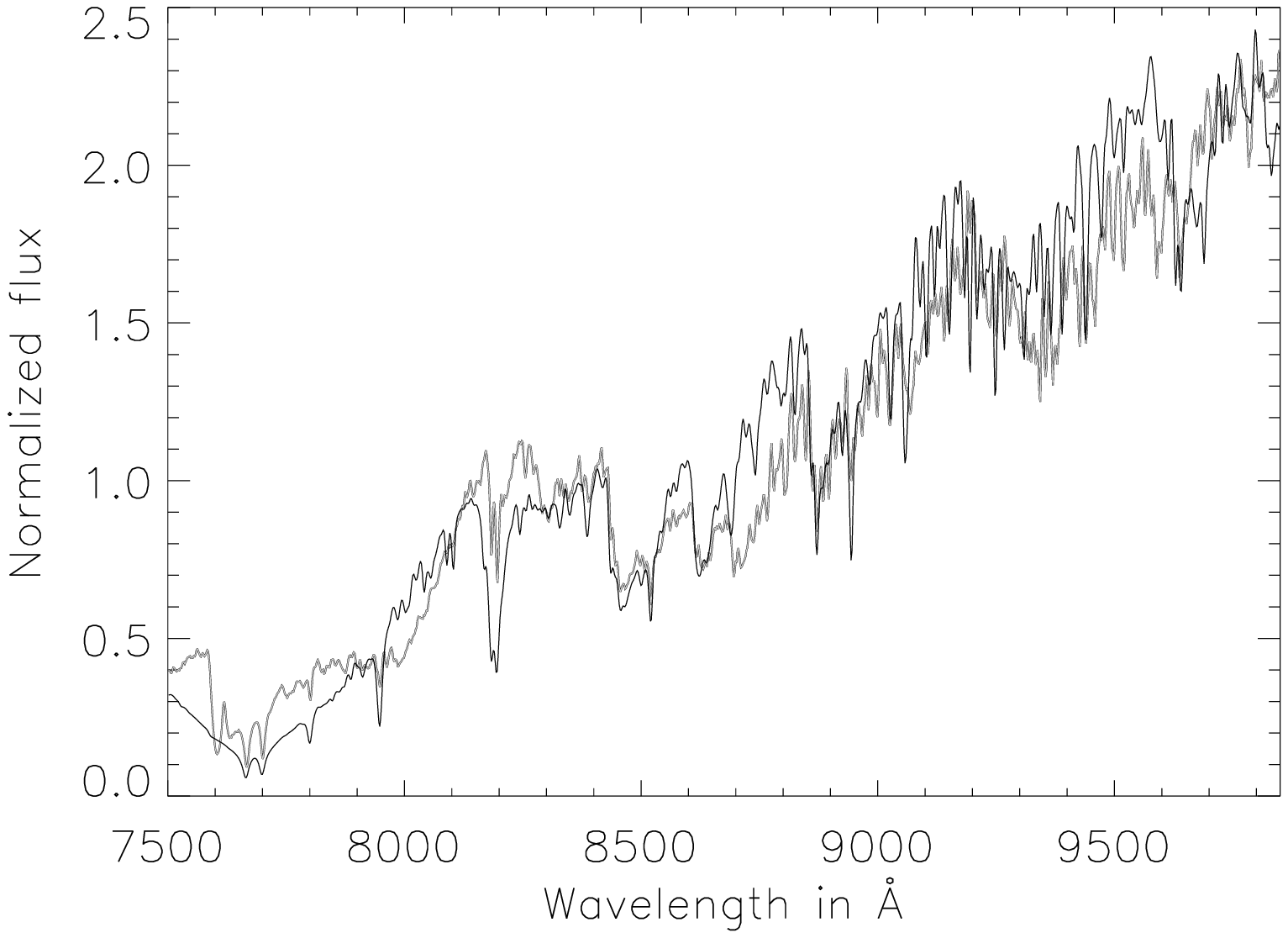}
\plotone{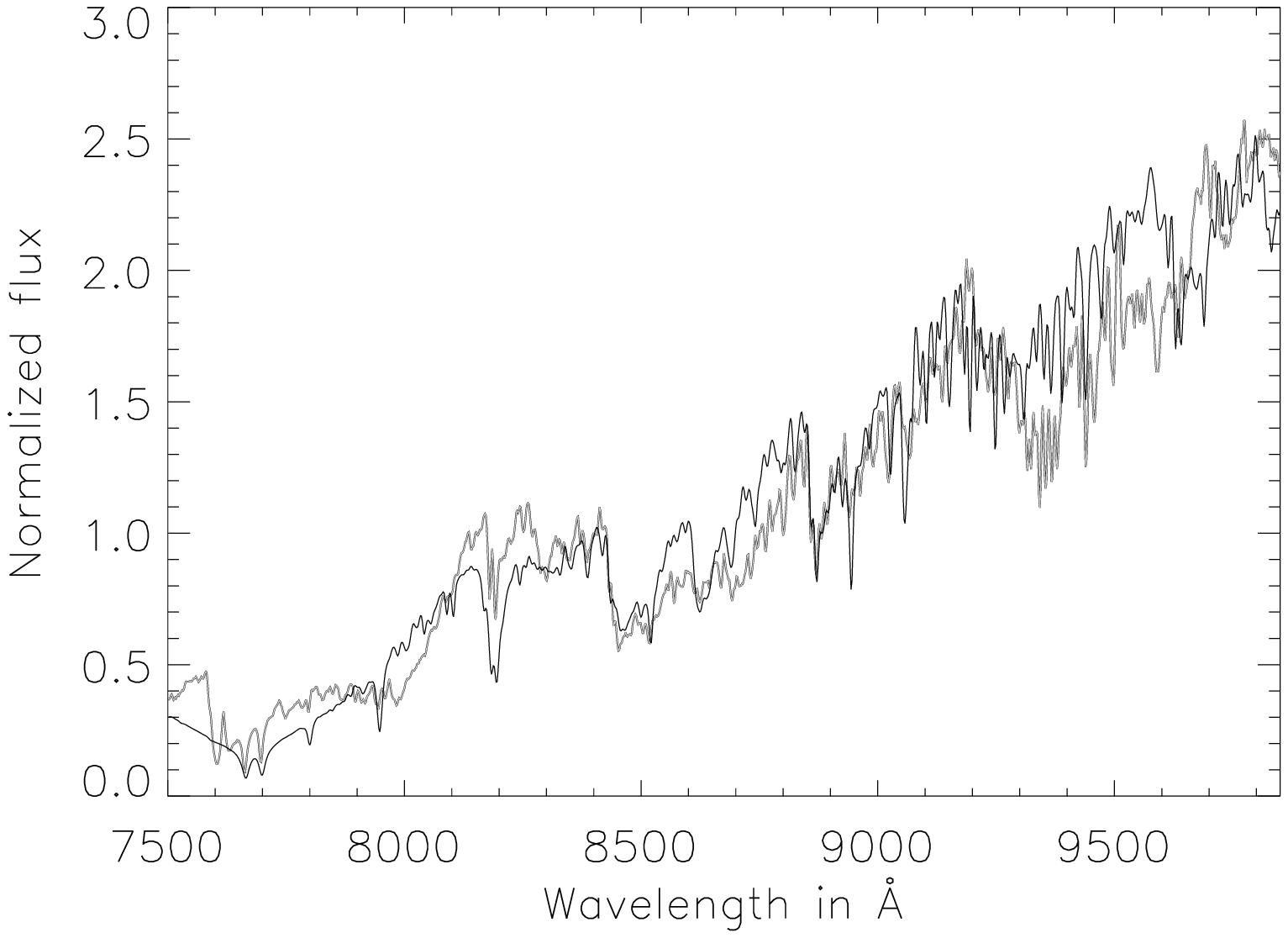}
\caption{\label{lowfitplot0149}
Fits (dark line) to the observed dwarfs (grey line) to
2M0149+2956 (left panel) and 2M2234+2359 (right panel).
All models are AMES-Dusty. See Tab. \ref{lowresfittab} for parameters.
}
\end{figure}

\begin{figure}
\epsscale{0.49}
\plotone{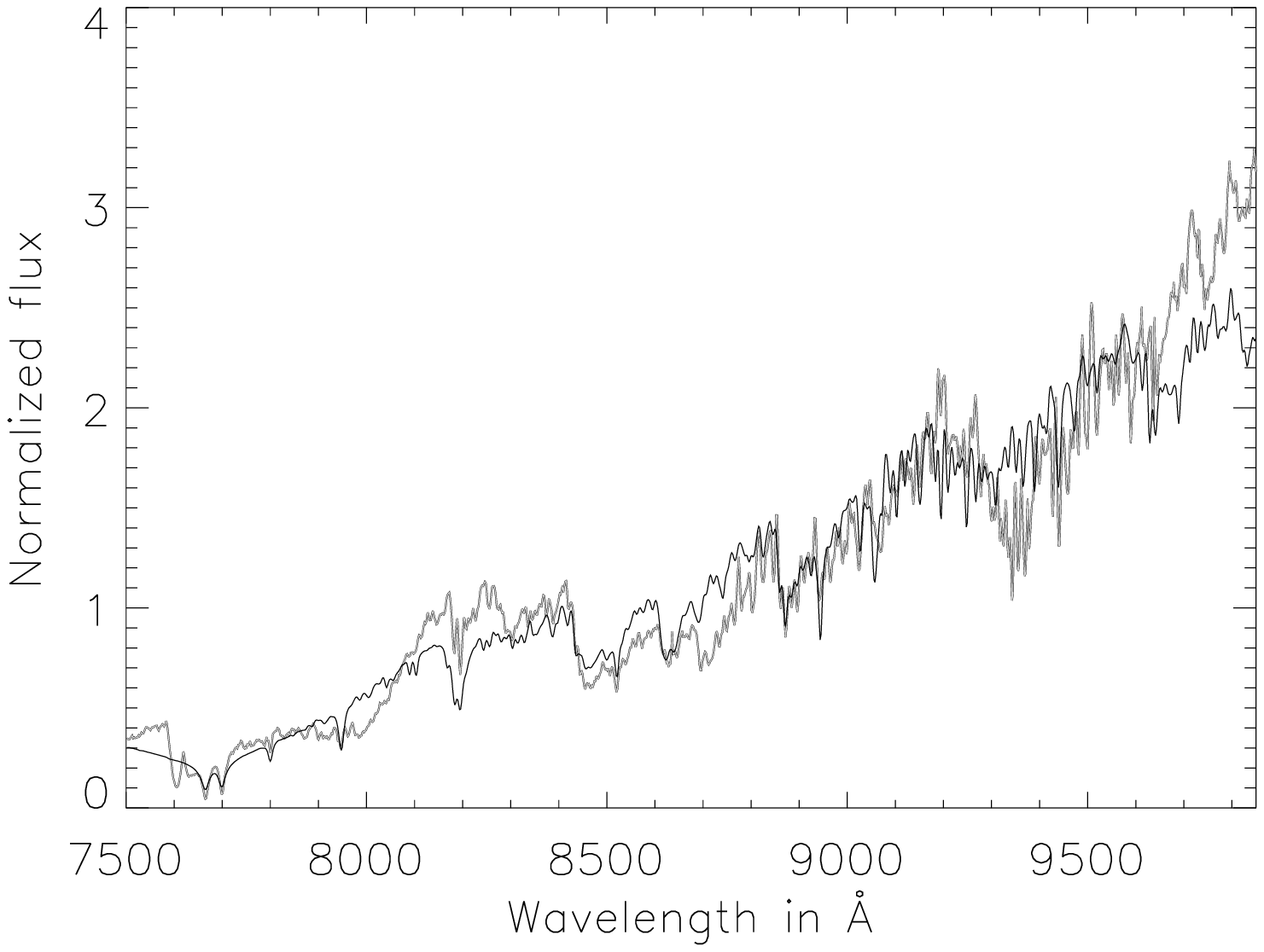}
\plotone{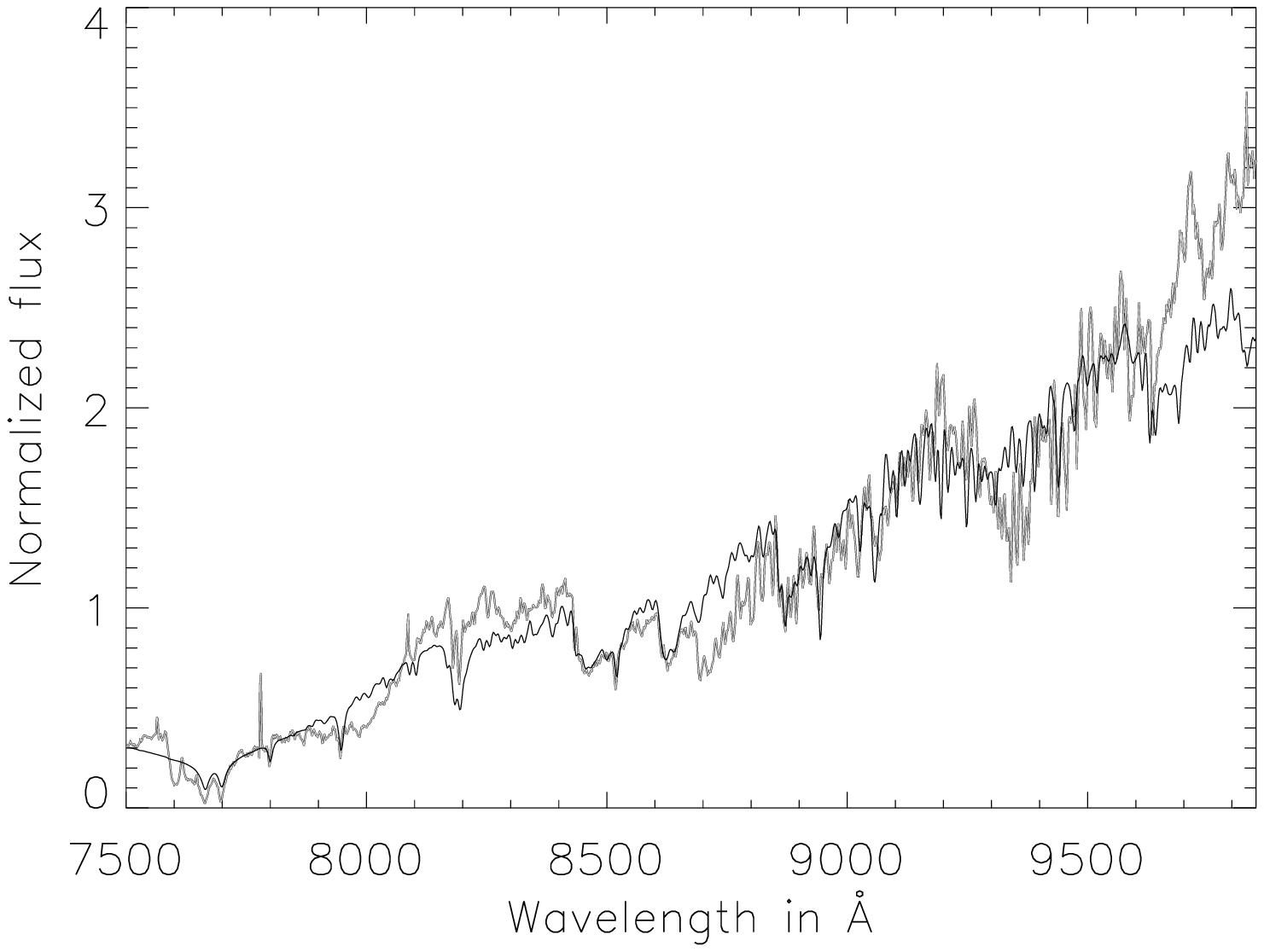}
\caption{\label{lowfitplot0345}
Fits (dark line) to the observed dwarfs (grey line) to
2M0345+2540 (left panel) and 2M0147+3453 (right panel).
All models are AMES-Dusty. See Tab. \ref{lowresfittab} for parameters.
}
\end{figure}

\begin{figure}
\epsscale{0.49}
\plotone{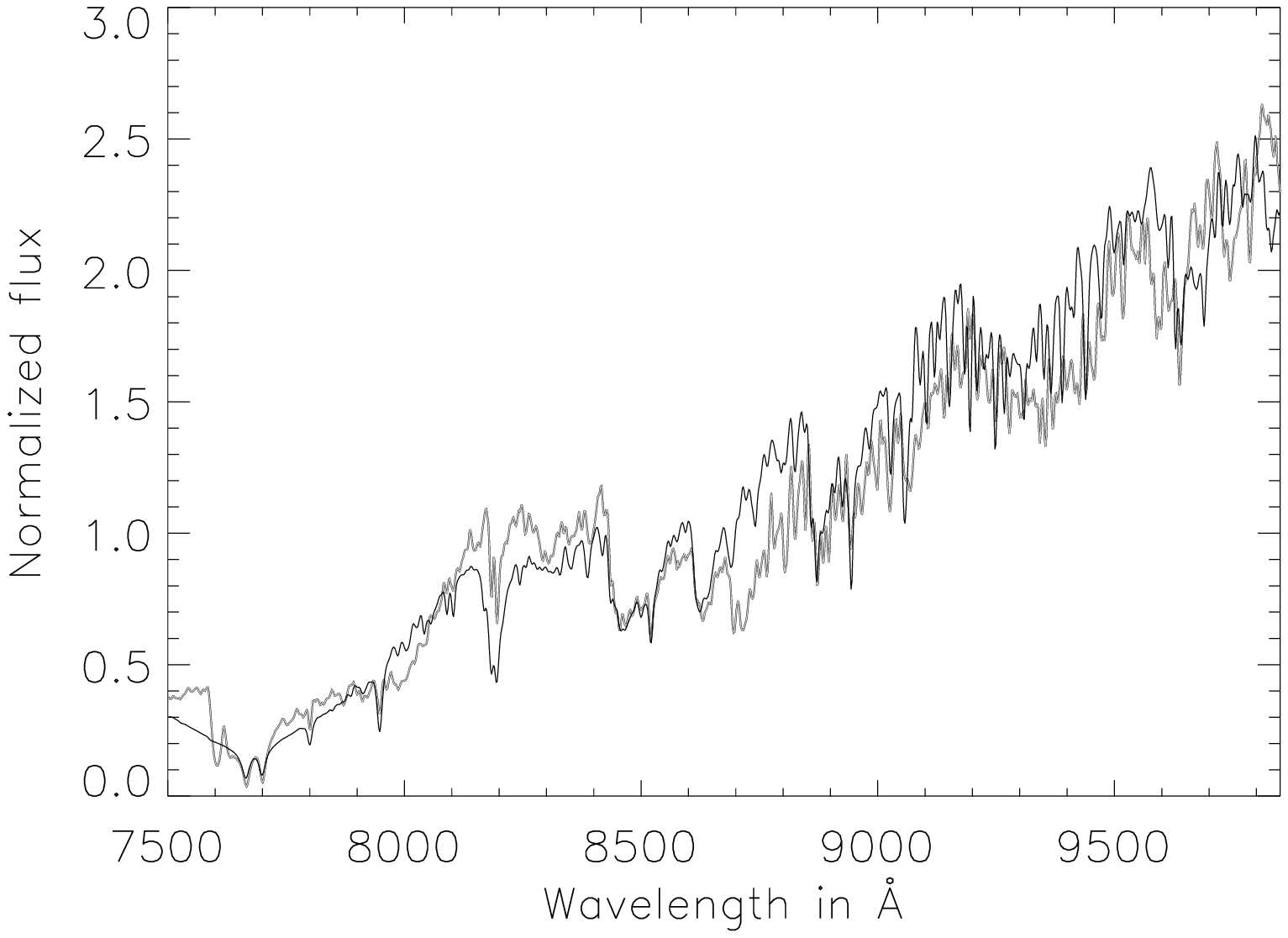}
\plotone{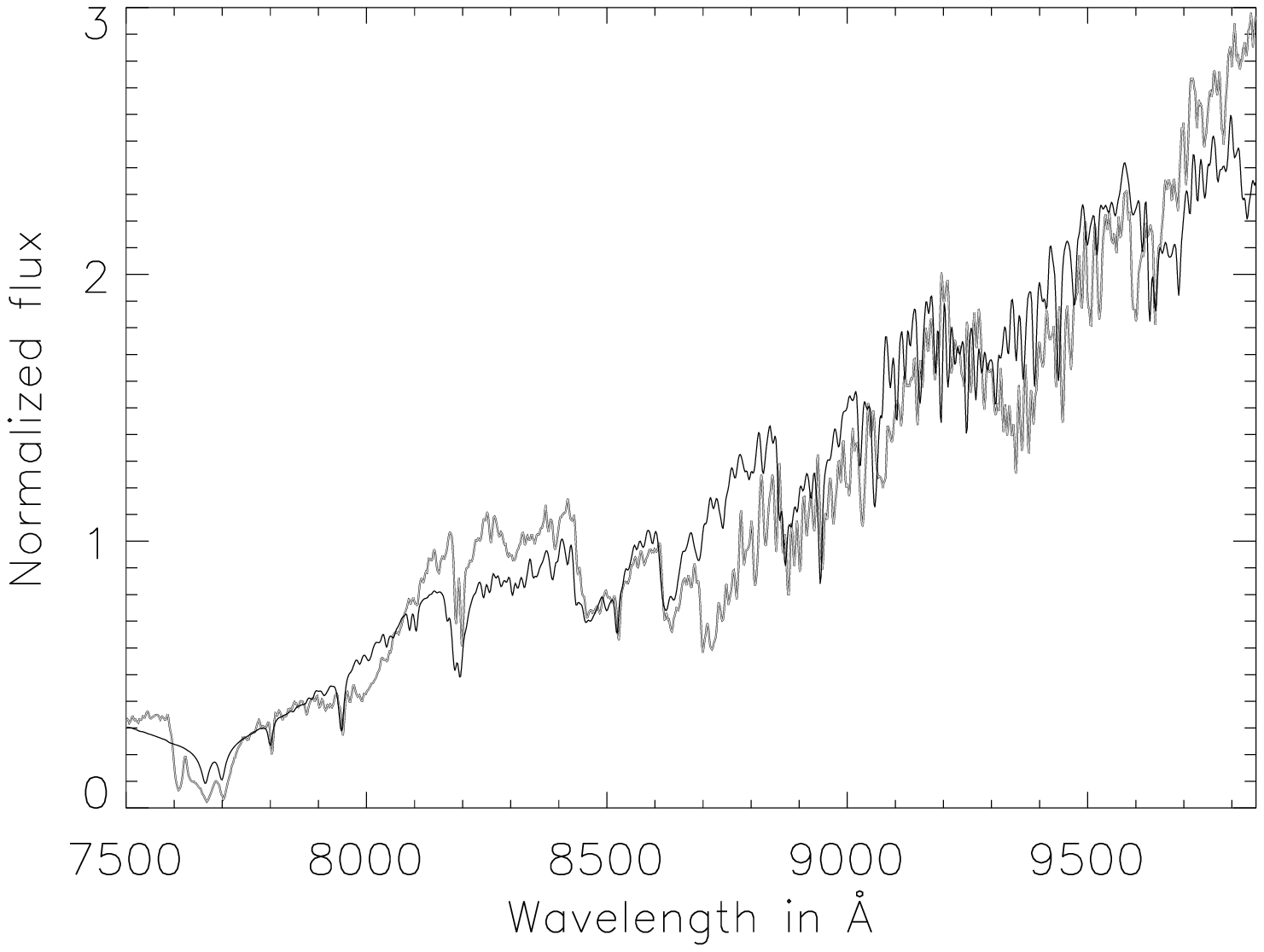}
\caption{\label{lowfitplot0746}
Fits (dark line) to the observed dwarfs (grey line) to
2M0746+200 (left panel) and 2M1439+1929 (right panel).
All models are AMES-Dusty. See Tab. \ref{lowresfittab} for parameters.
}
\end{figure}

\begin{figure}
\epsscale{0.49}
\plotone{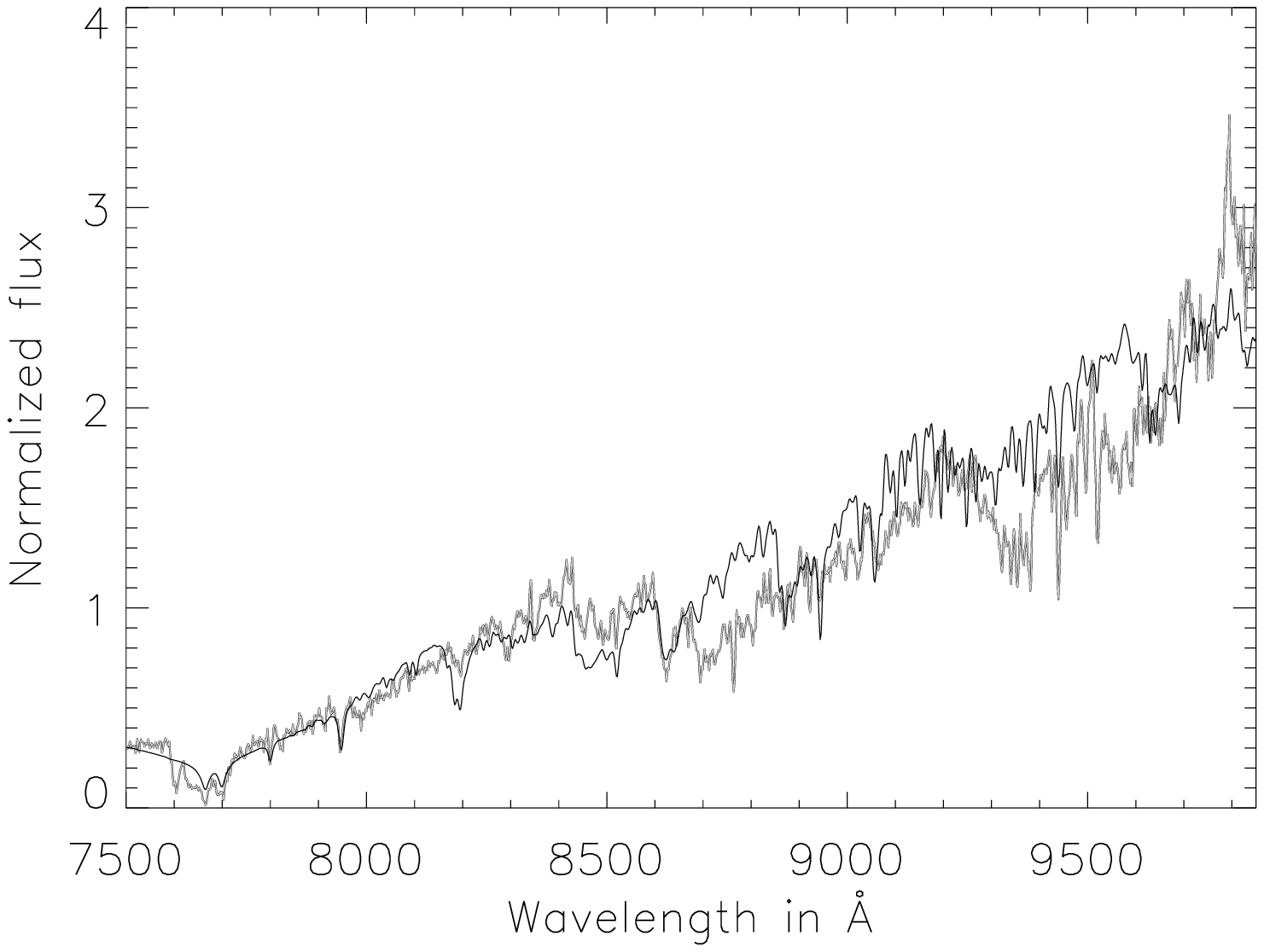}
\plotone{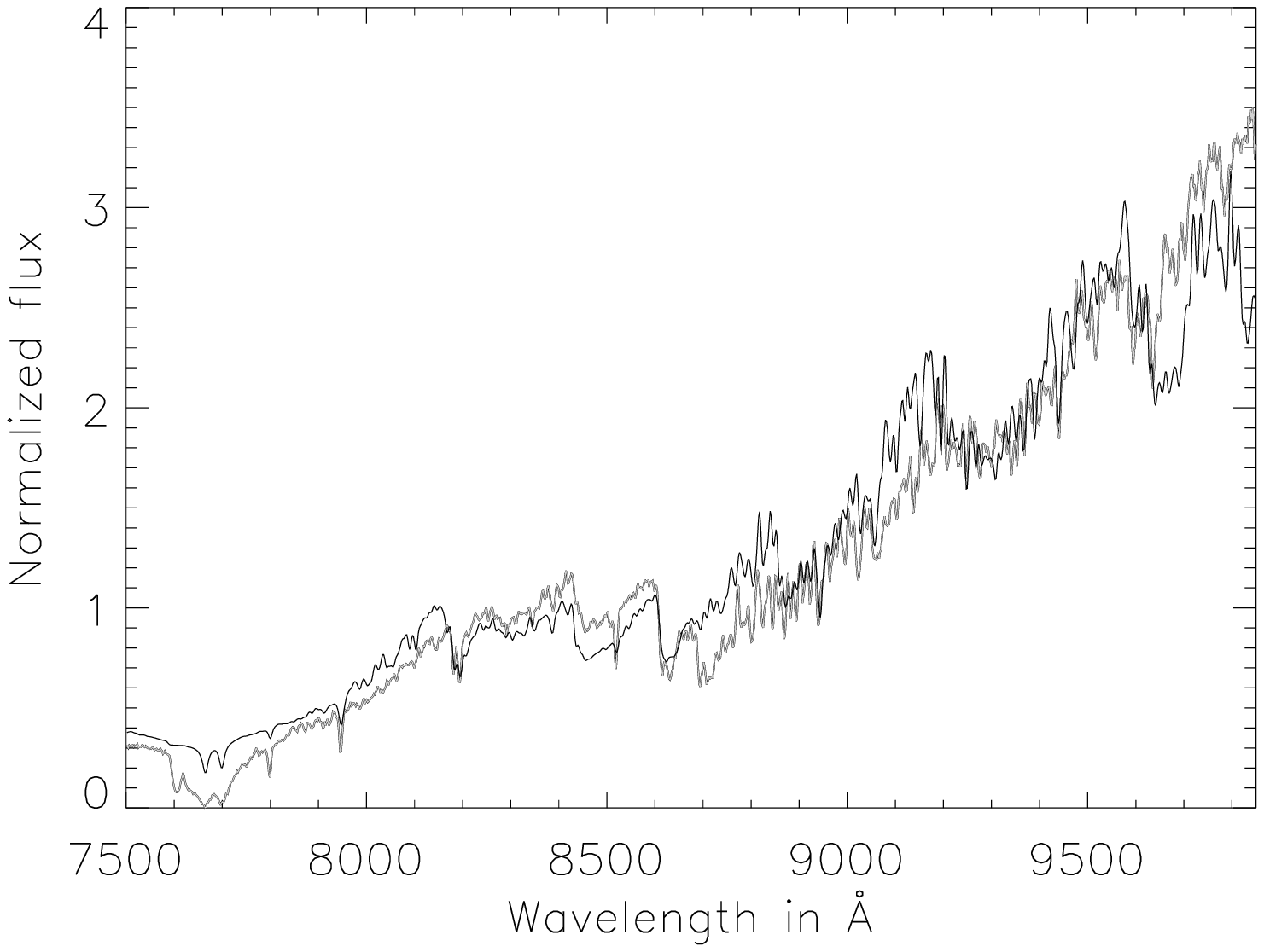}
\caption{\label{lowfitplot1726}
Fits (dark line) to the observed dwarfs (grey line) to
2M1726+1538 (left panel) and 2M1146+2230 (right panel).
All models are AMES-Dusty. See Tab. \ref{lowresfittab} for parameters.
}
\end{figure}

\begin{figure}
\epsscale{0.49}
\plotone{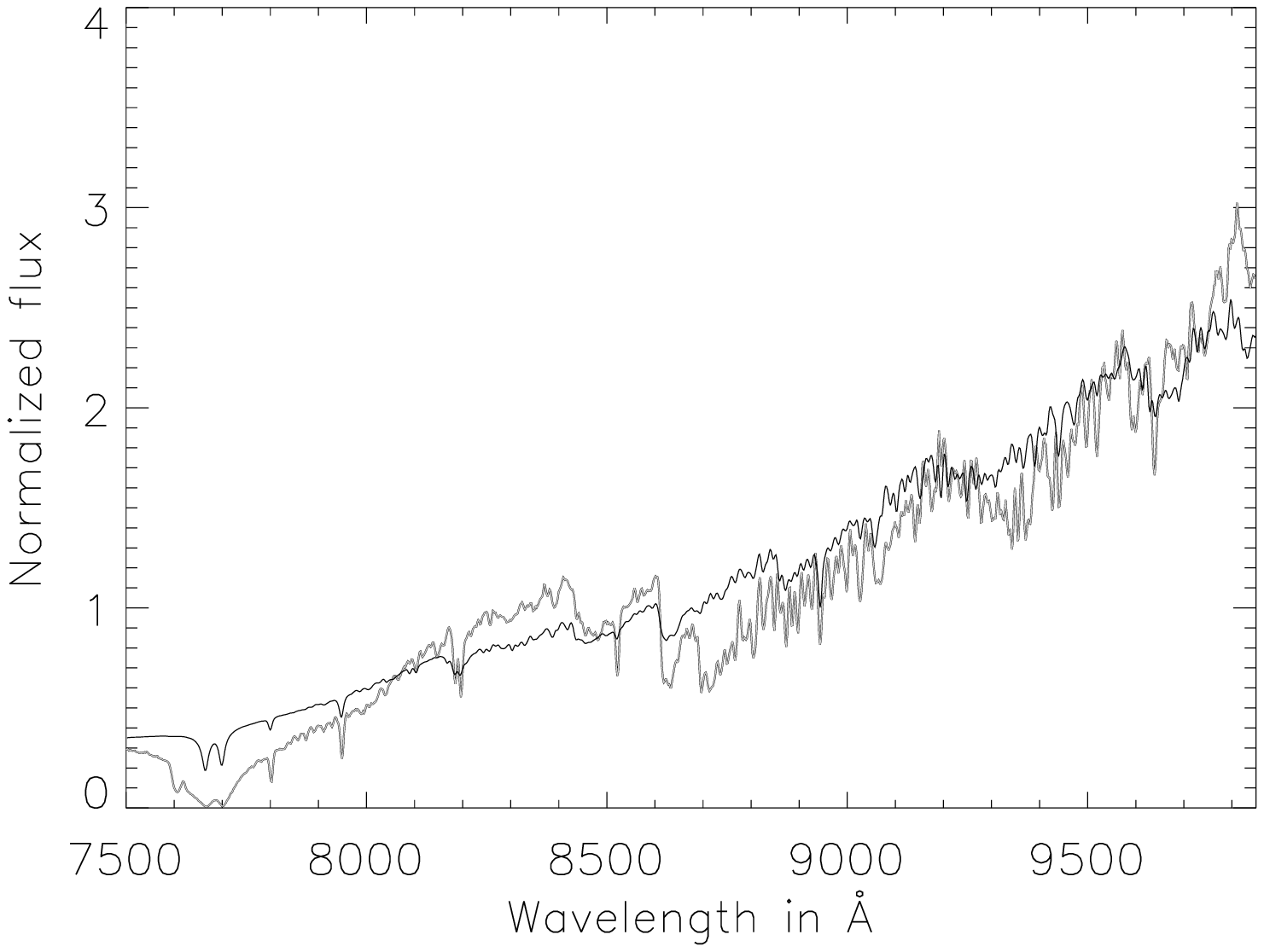}
\plotone{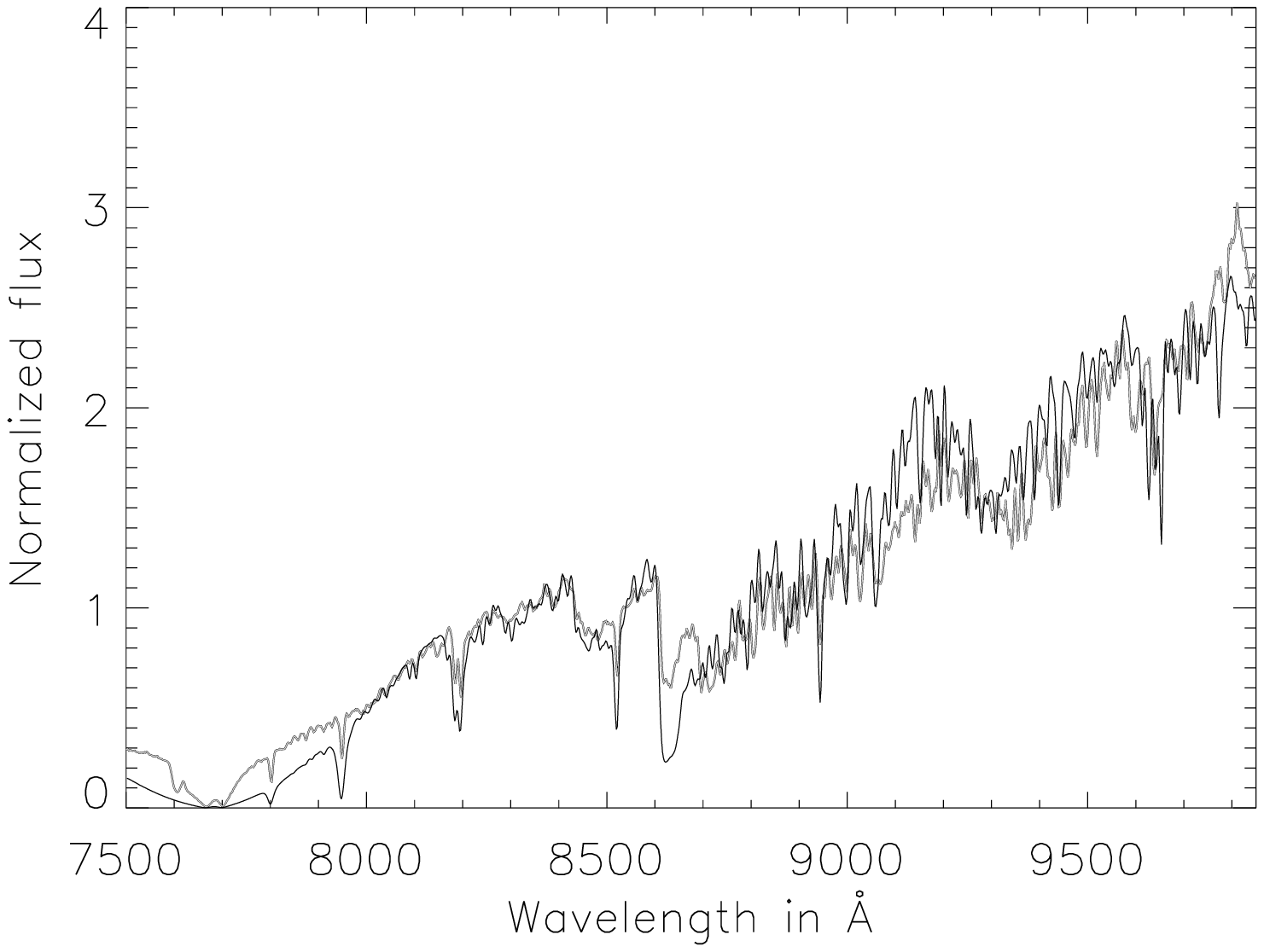}
2M0036+1821\\
\caption{\label{lowfitplot0036}
Fits (dark line) to the observed dwarfs (grey line) to
2M0036+1821.
The model in the left panel is AMES-Dusty, the model in the left panel is AMES-Cond.
See Tab. \ref{lowresfittab} for parameters.
}
\end{figure}

\begin{figure}
\epsscale{0.49}
\plotone
{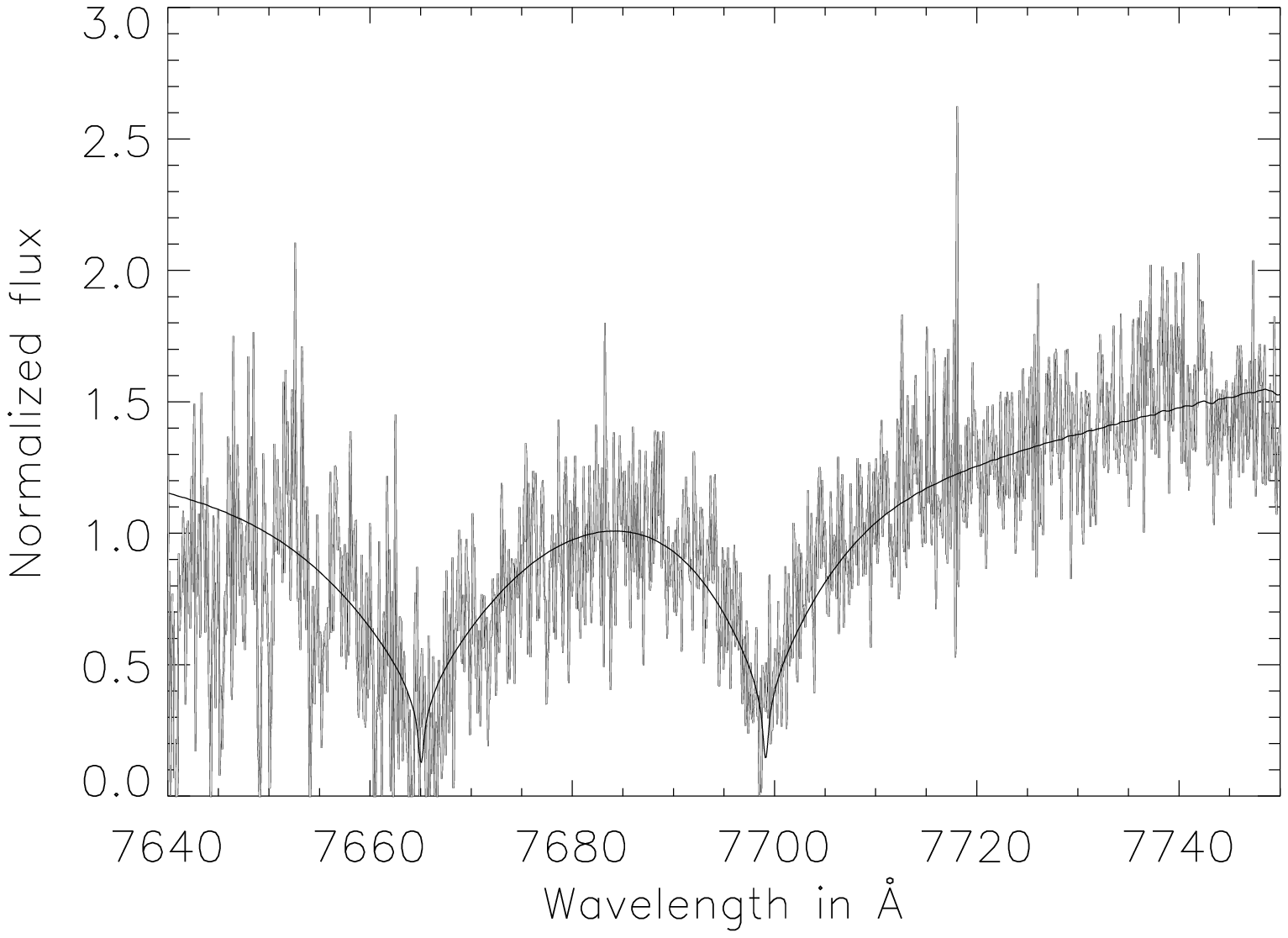}
\plotone
{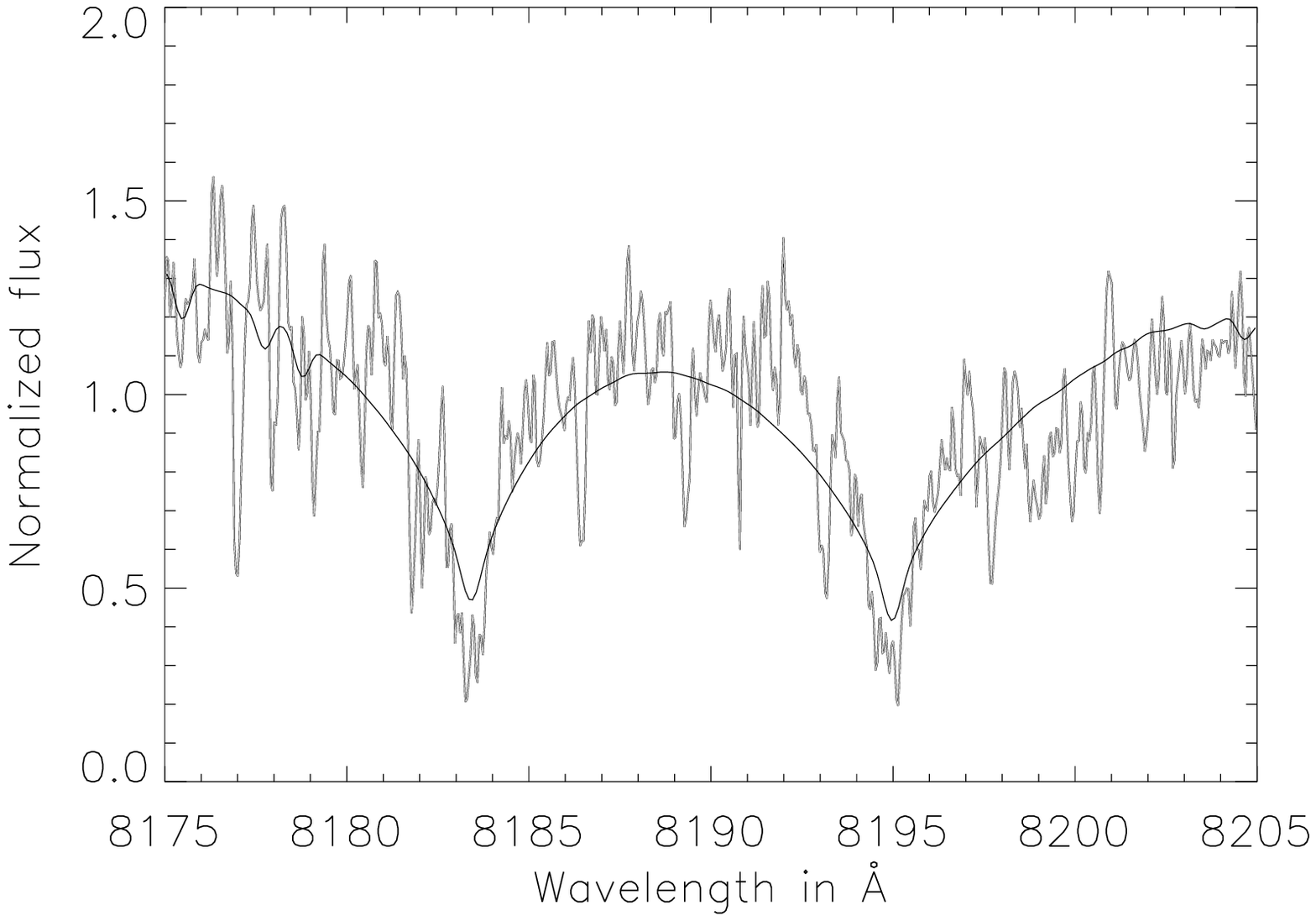}\\
\plotone
{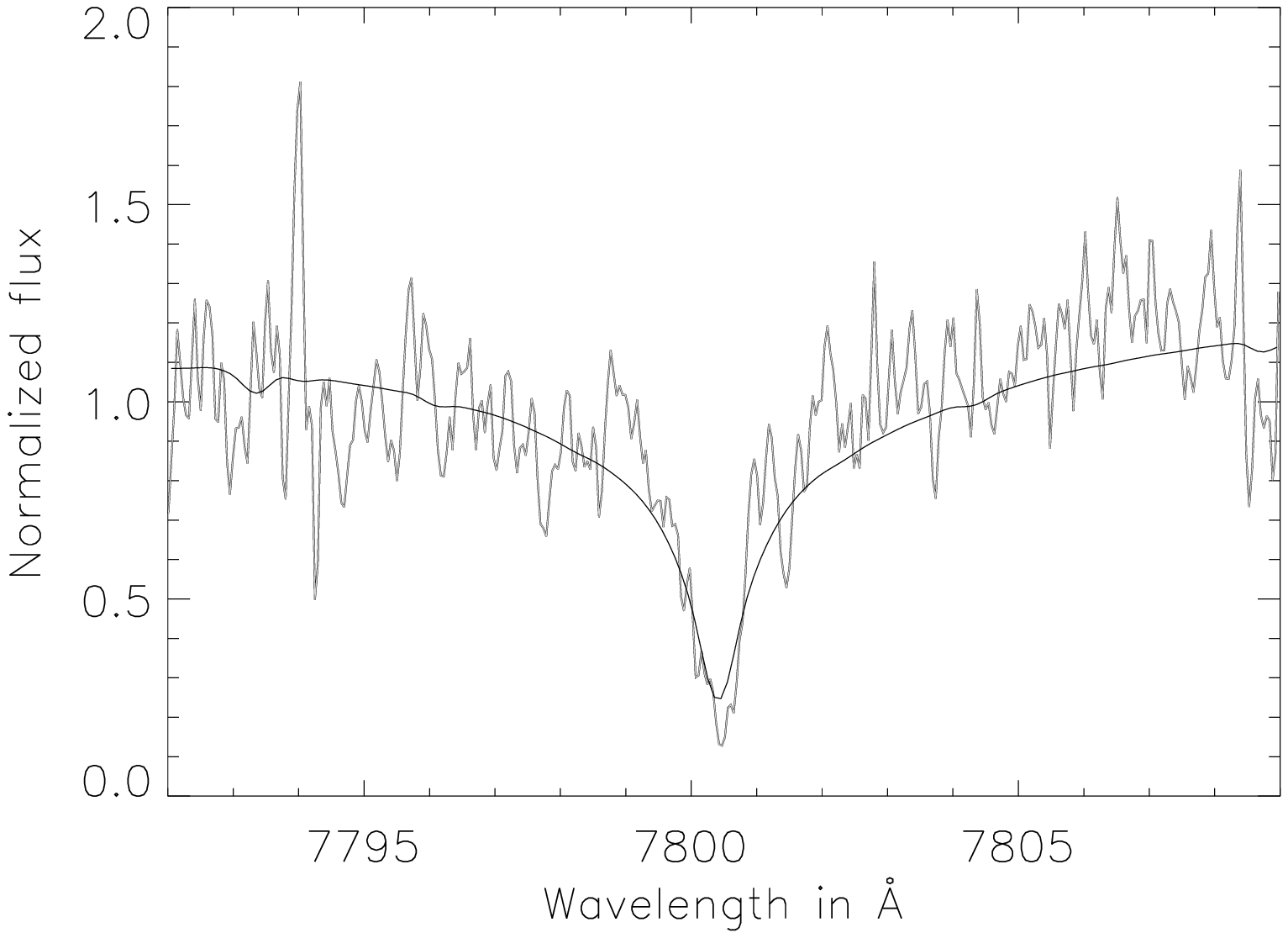}
\plotone
{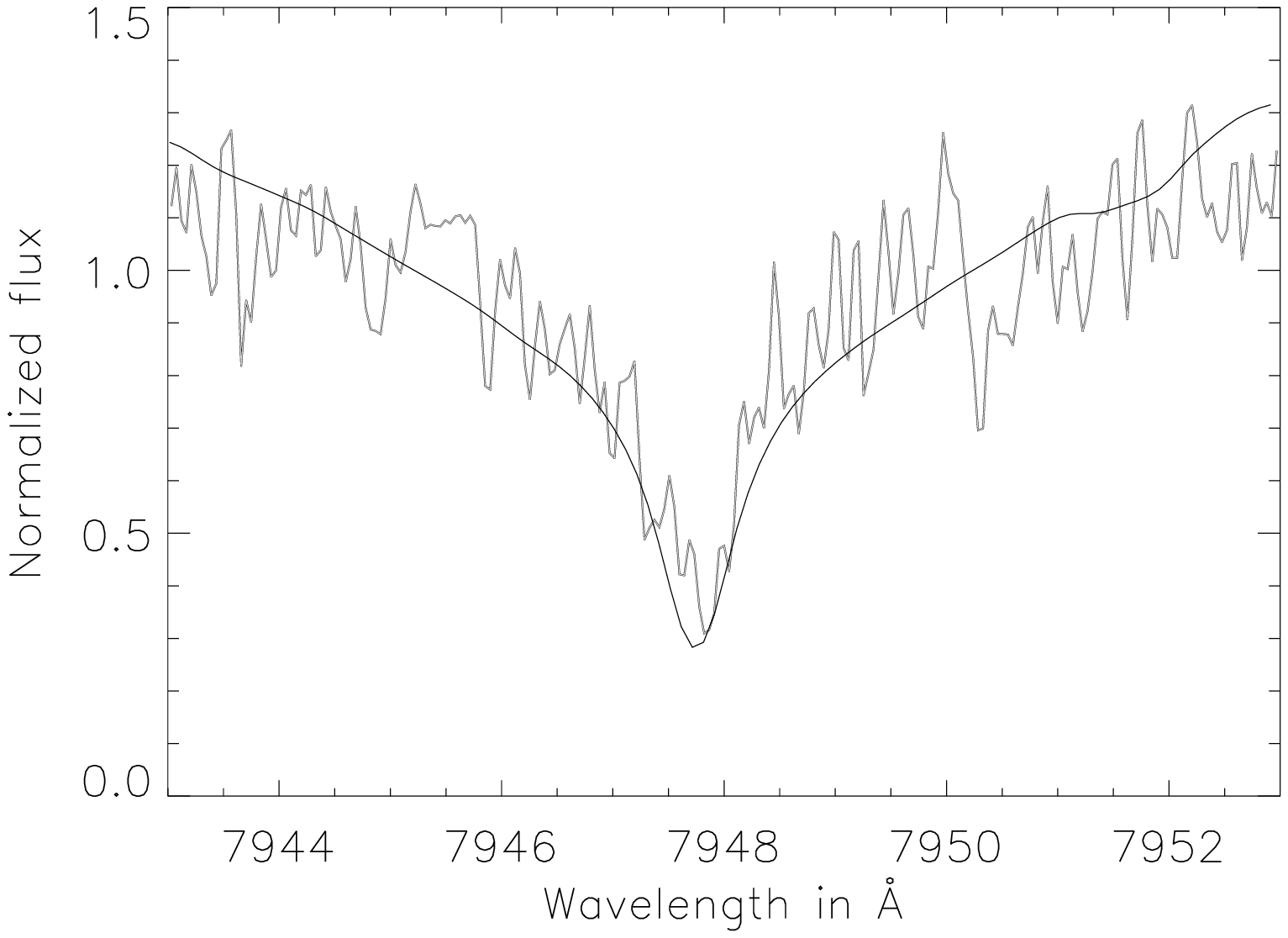}\\
\plotone
{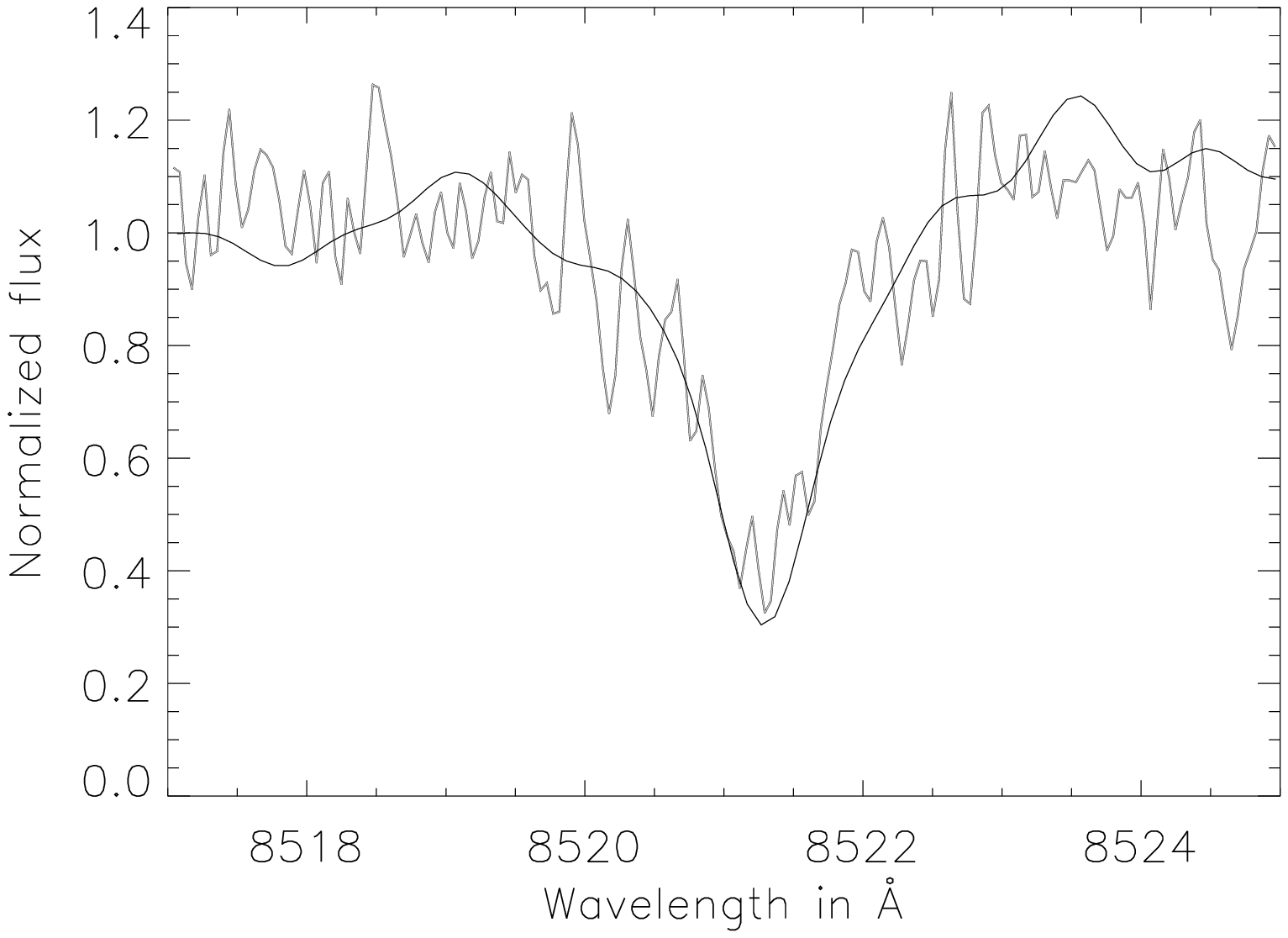}
\caption{\label{hi0149fit}
Fits (dark line) to 2M0149+2956 (grey line).
See Tab. \ref{highresfittab} for parameters.
Telluric features have not been removed.
}
\end{figure}

\begin{figure}
\epsscale{0.49}
\plotone
{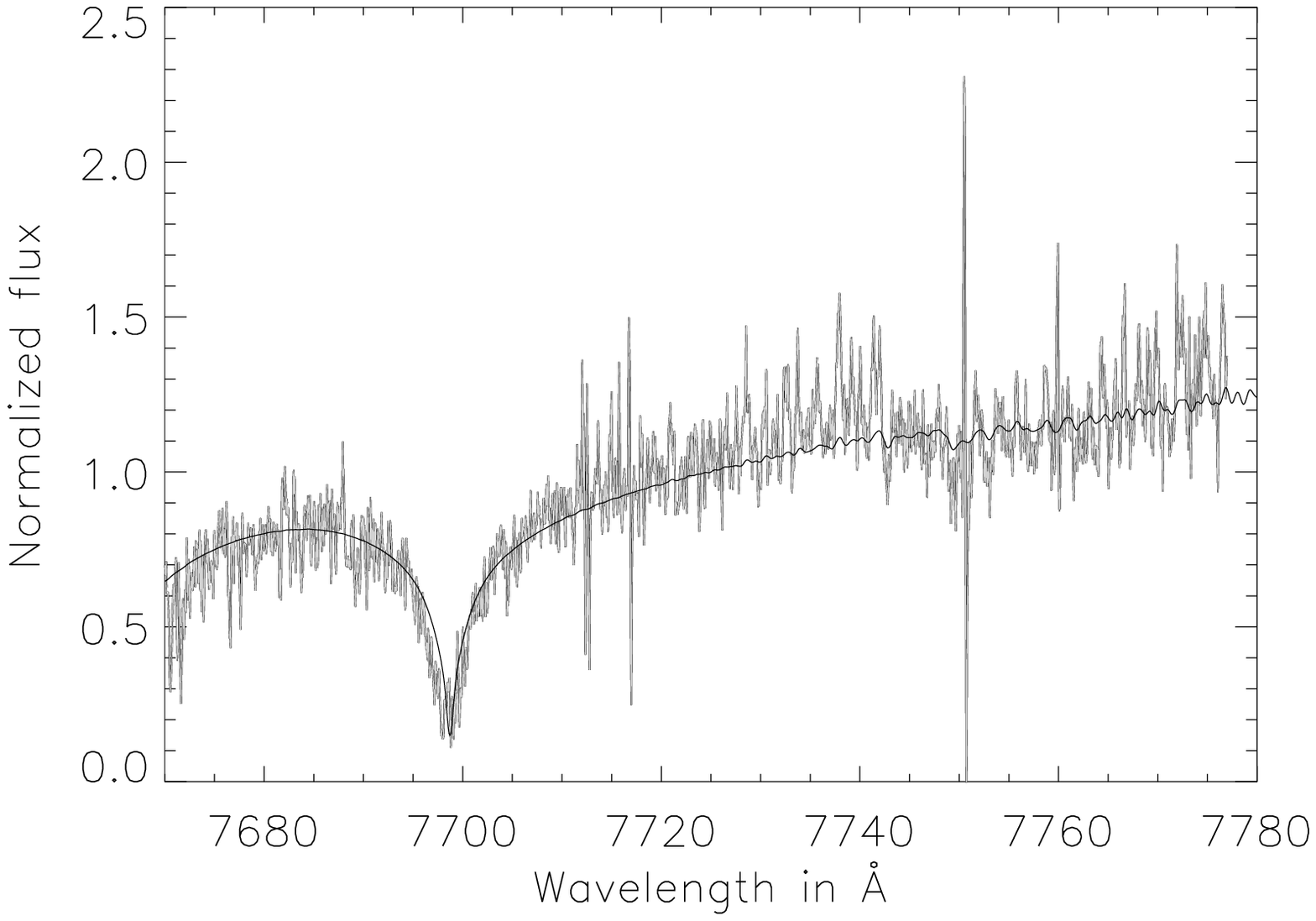}
\caption{\label{hi2234fit}
Fits (dark line) to 2M2234+2359 (grey line).
See Tab. \ref{highresfittab} for parameters.
Telluric features have not been removed.
}
\end{figure}

\begin{figure}
\epsscale{0.49}
\plotone
{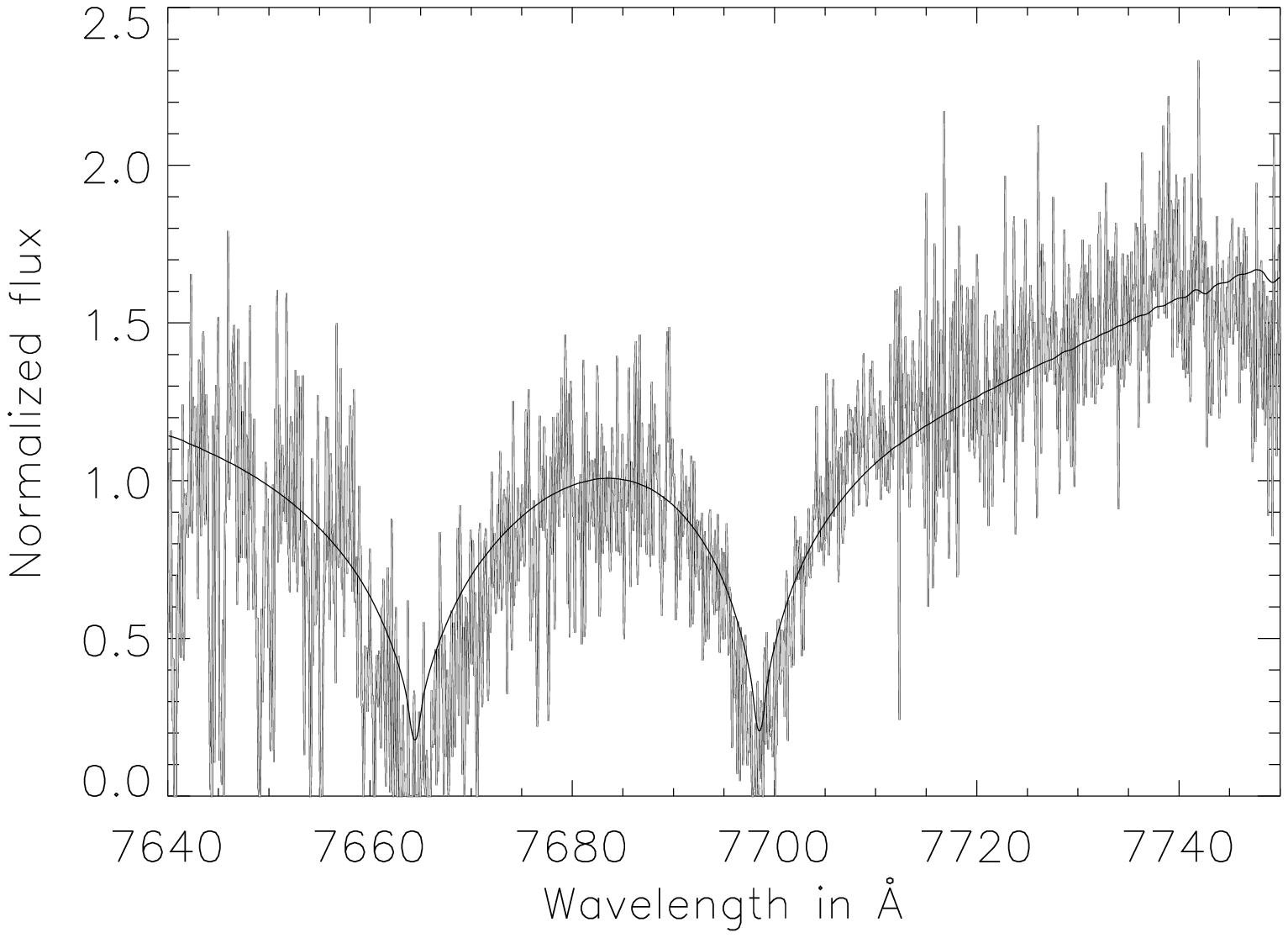}
\plotone
{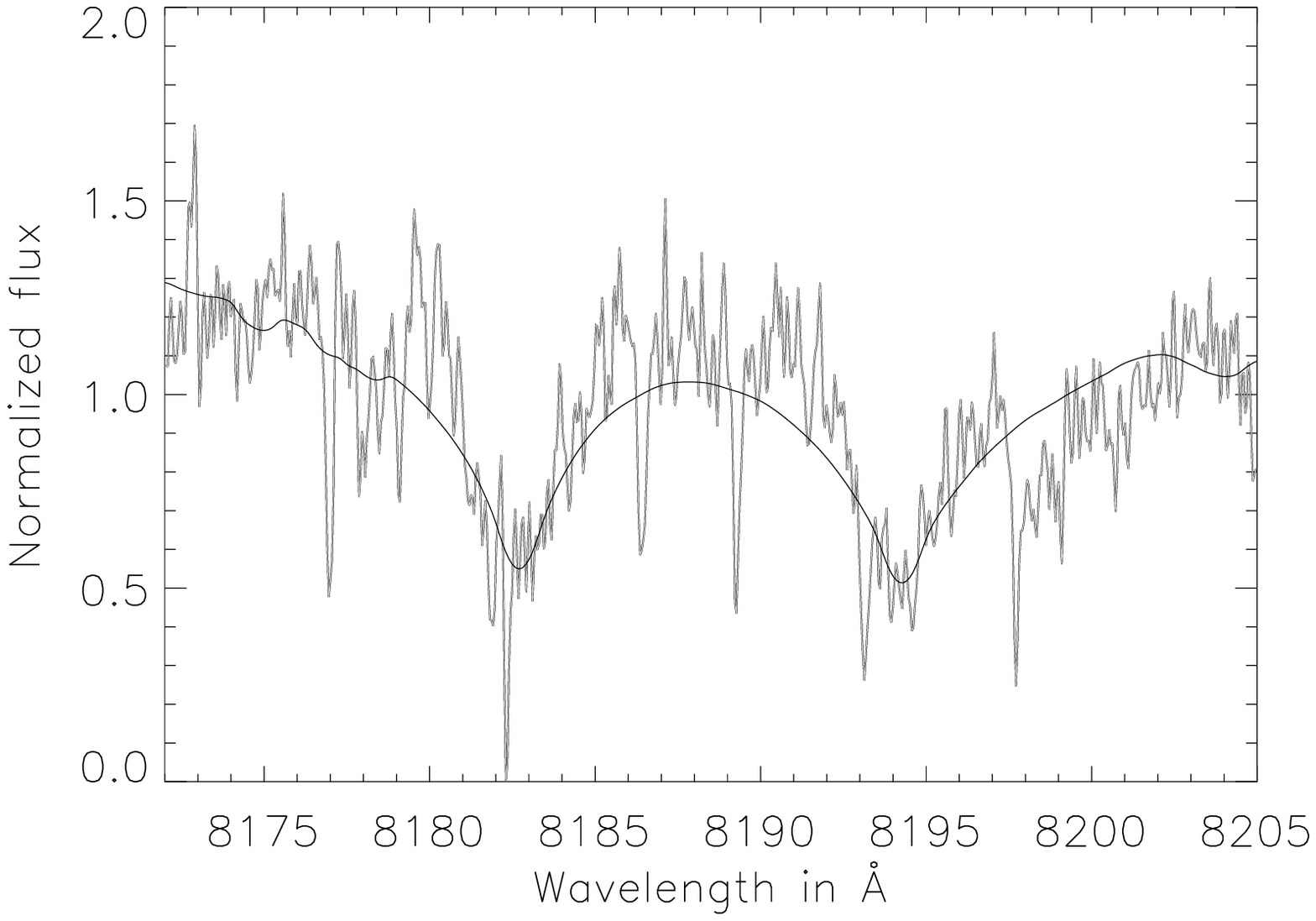}\\
\plotone
{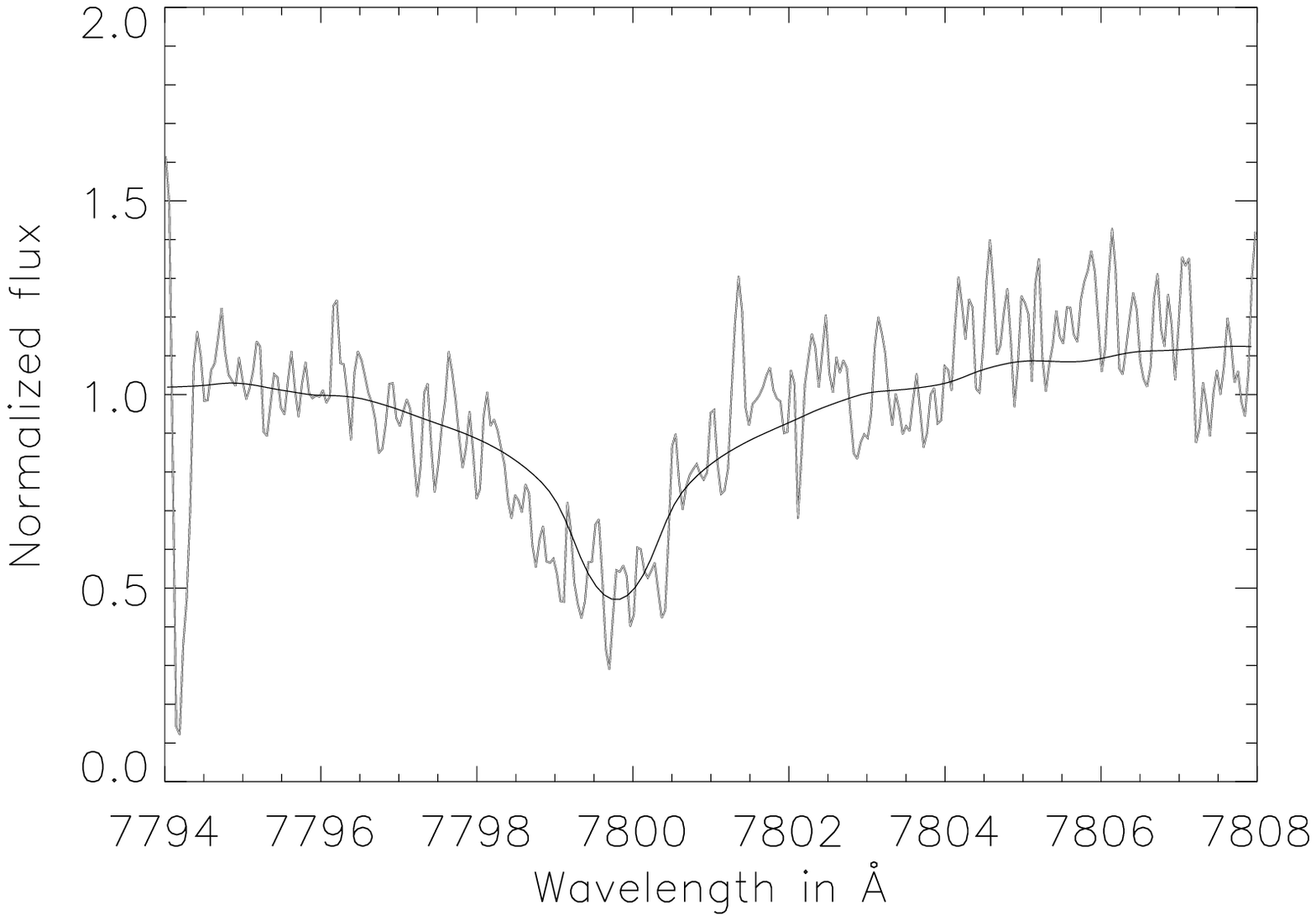}
\plotone
{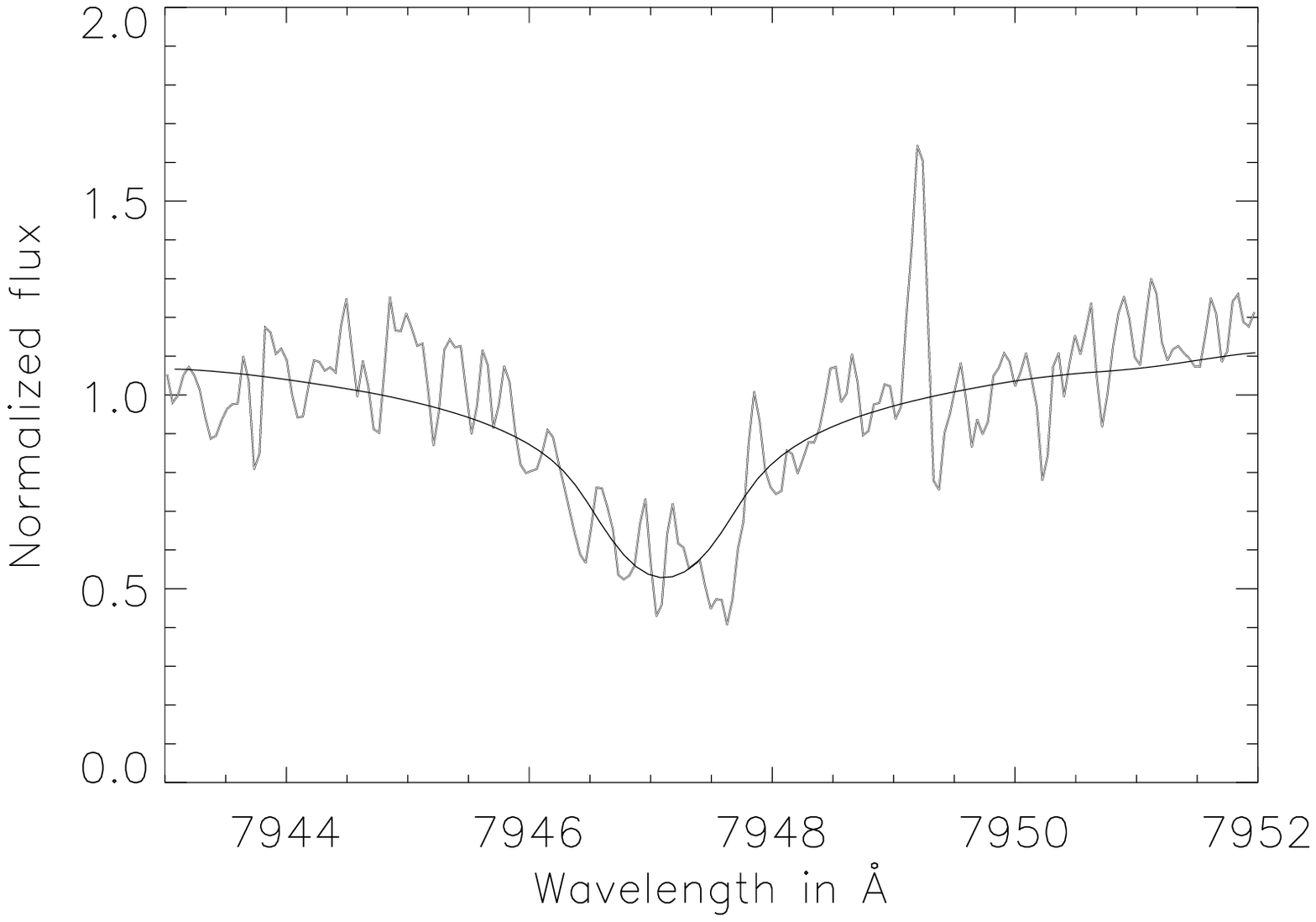}\\
\plotone
{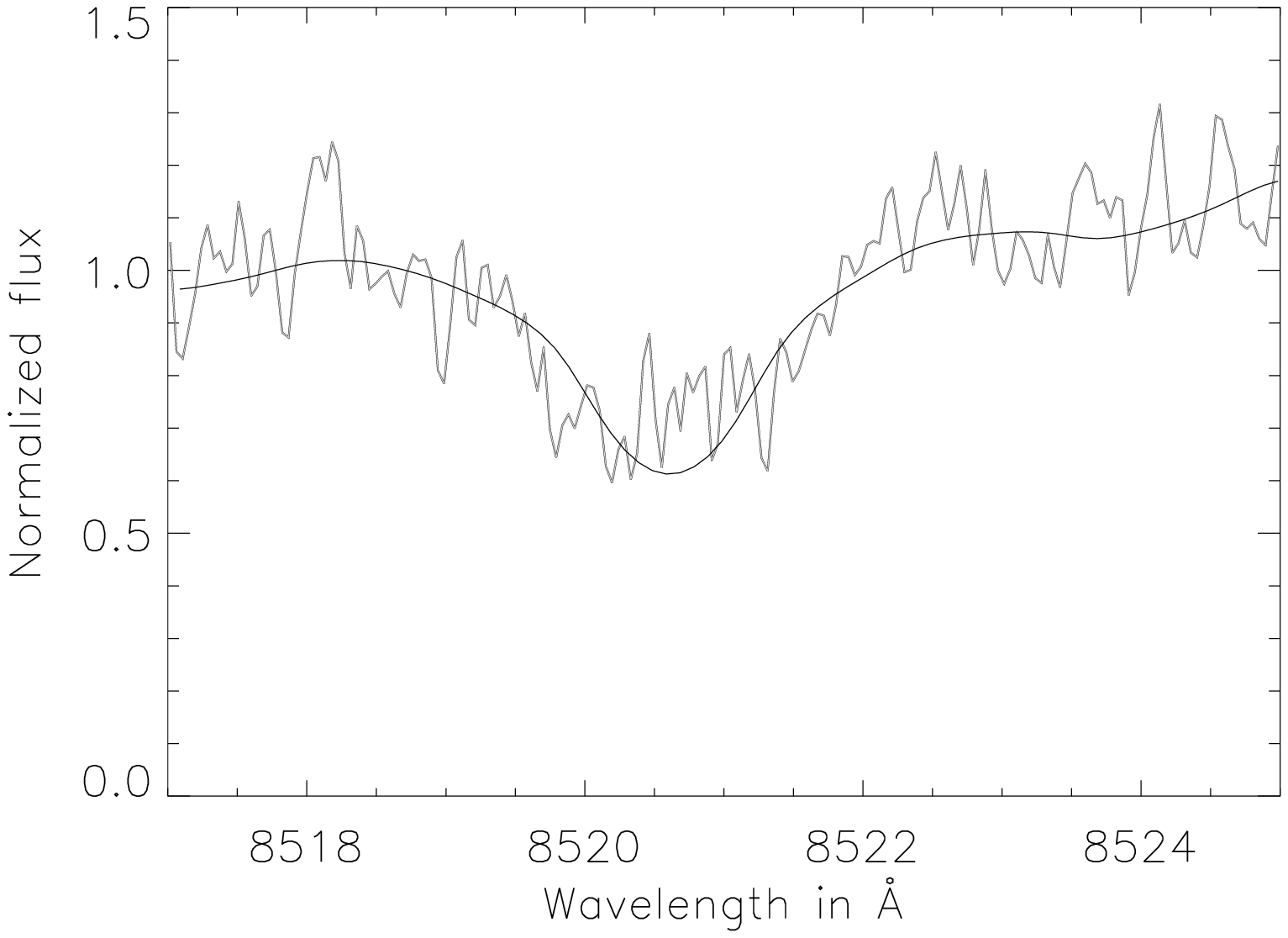}
\caption{\label{hi0345fit}
Fits (dark line) to 2M0345+2540 (grey line).
See Tab. \ref{highresfittab} for parameters.
Telluric features have not been removed.
}
\end{figure}

\begin{figure}
\epsscale{0.49}
\plotone
{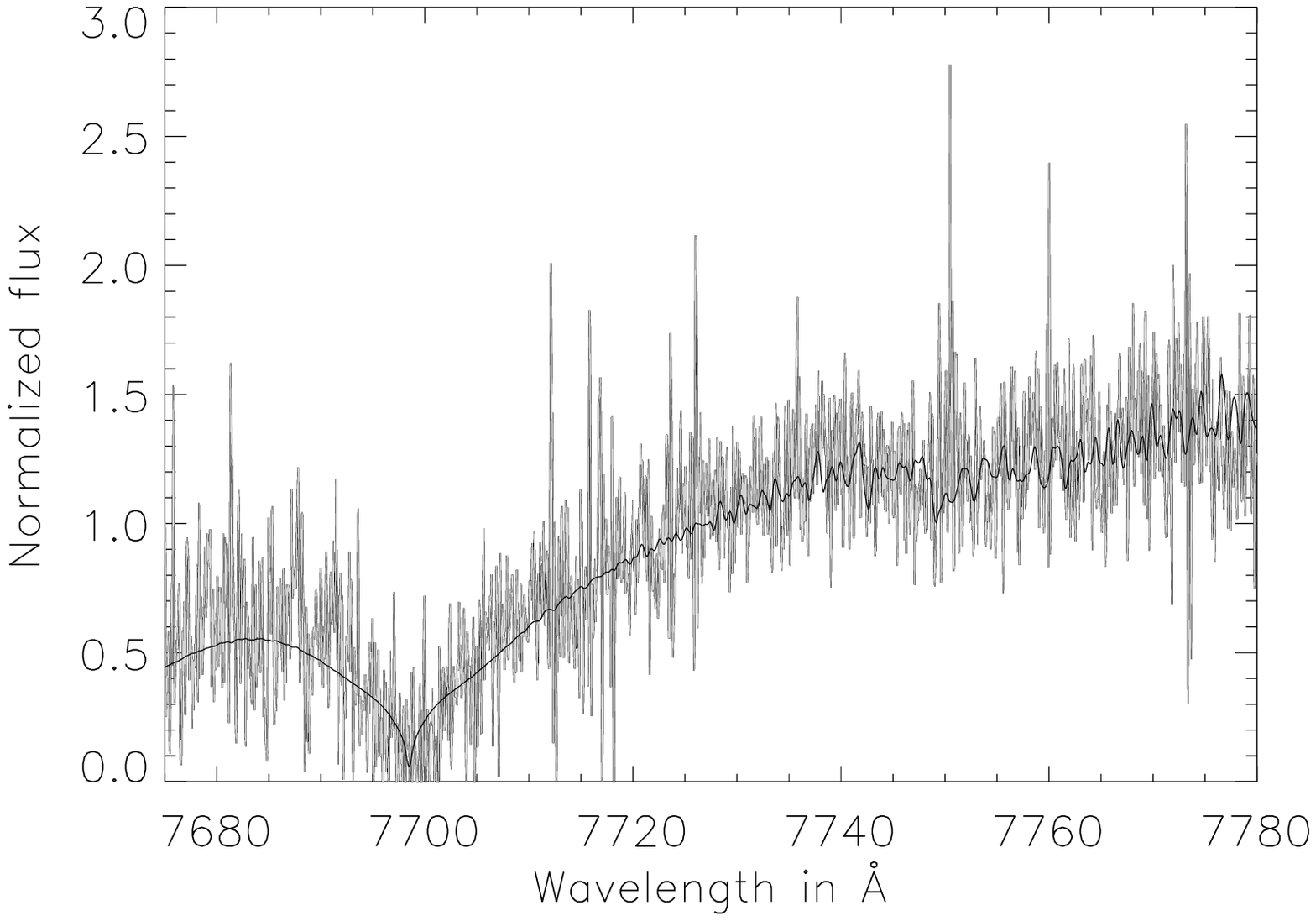}
\plotone
{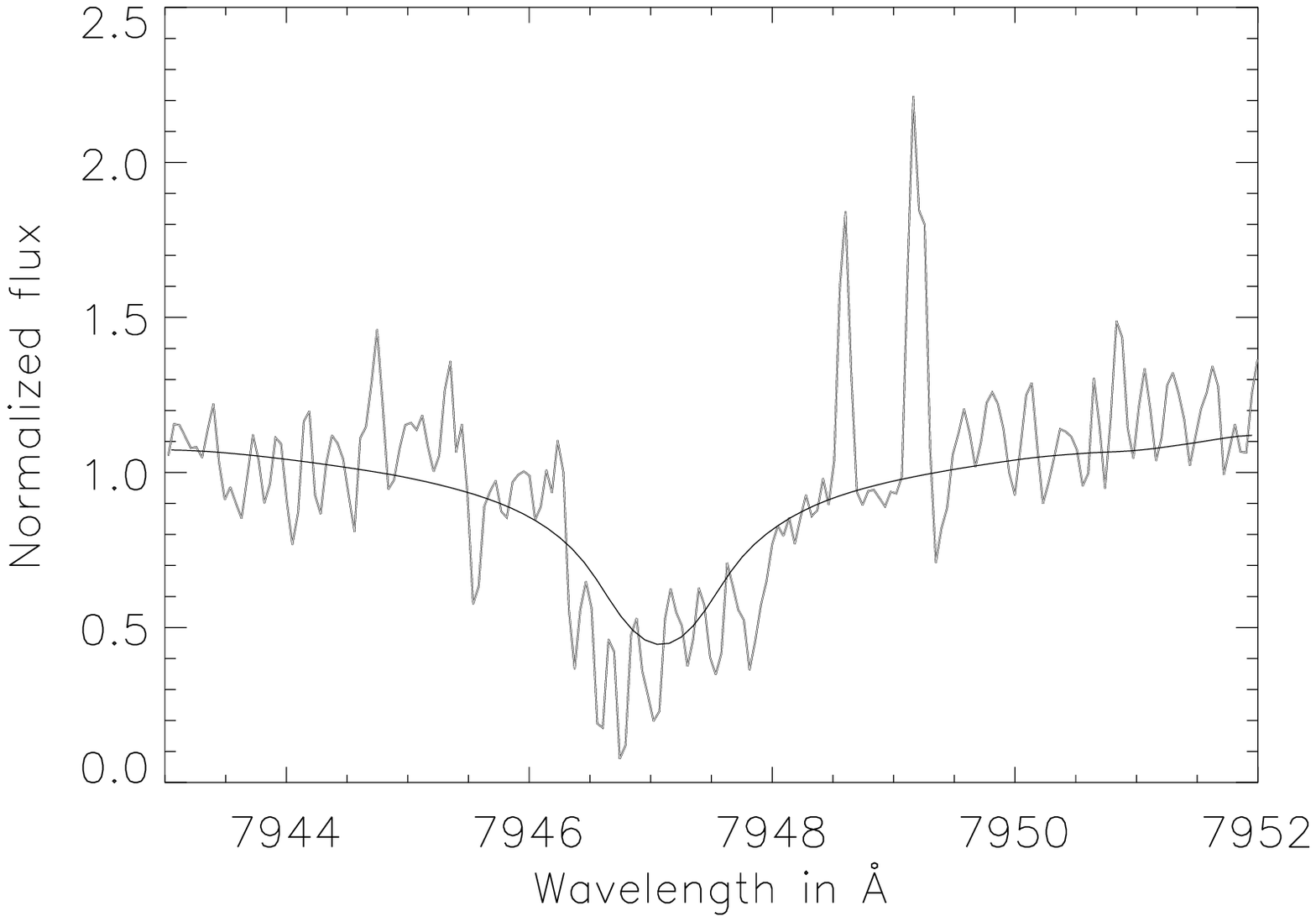}\\
\plotone
{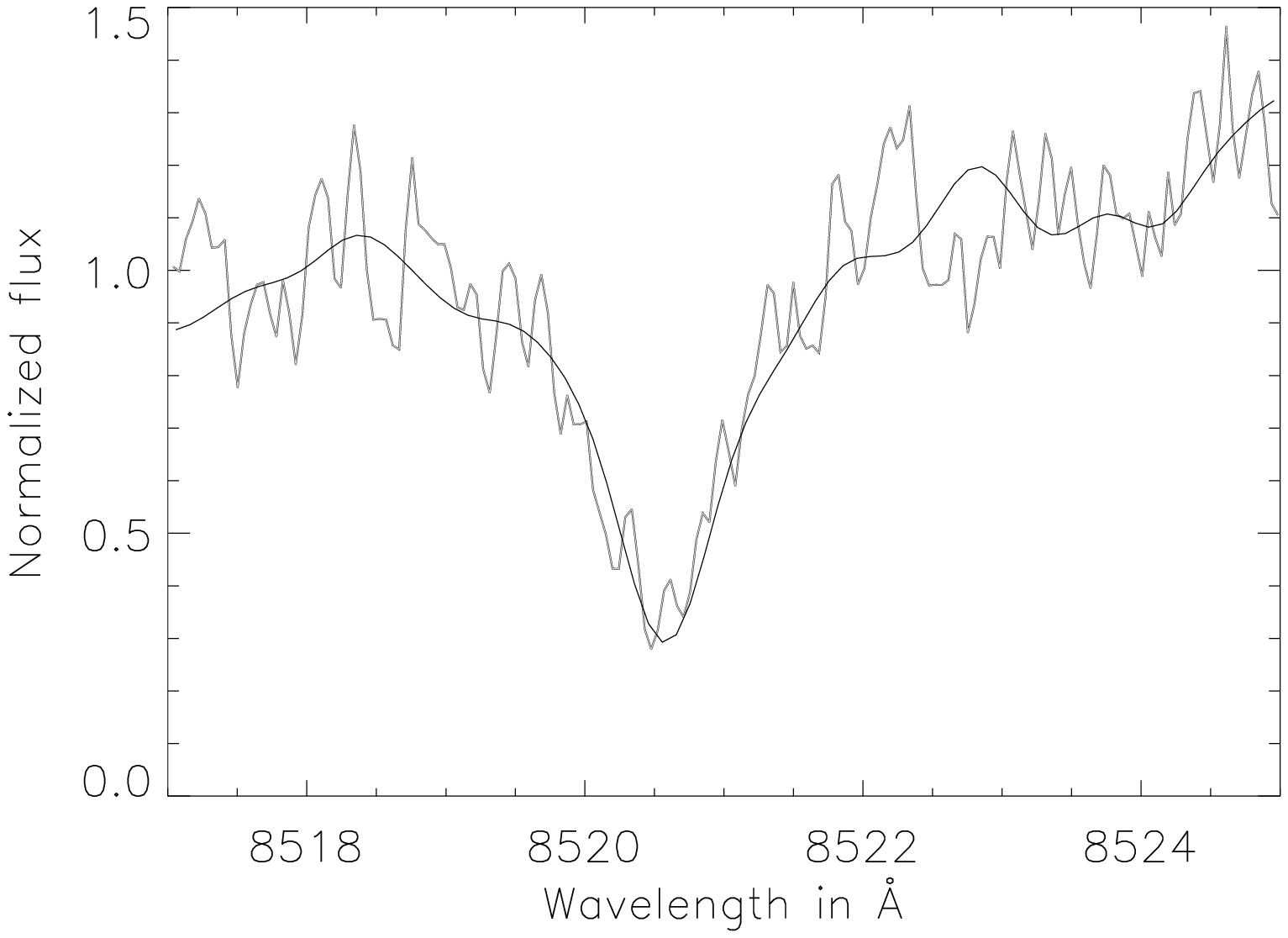}
\caption{\label{hi0147fit}
Fits (dark line) to 2M0147+3453 (grey line).
See Tab. \ref{highresfittab} for parameters.
Telluric features have not been removed.
}
\end{figure}

\begin{figure}
\epsscale{0.49}
\plotone
{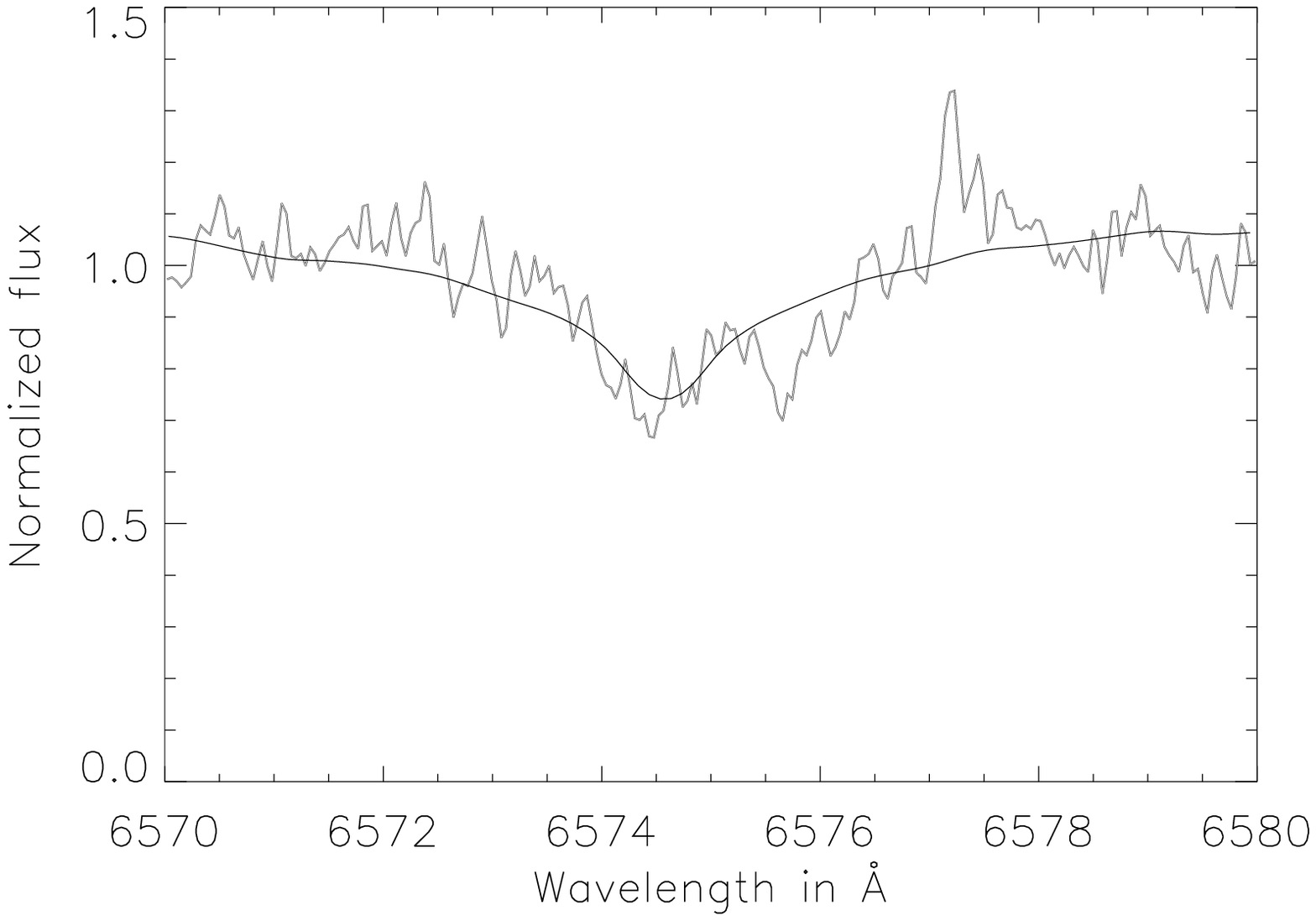}
\plotone
{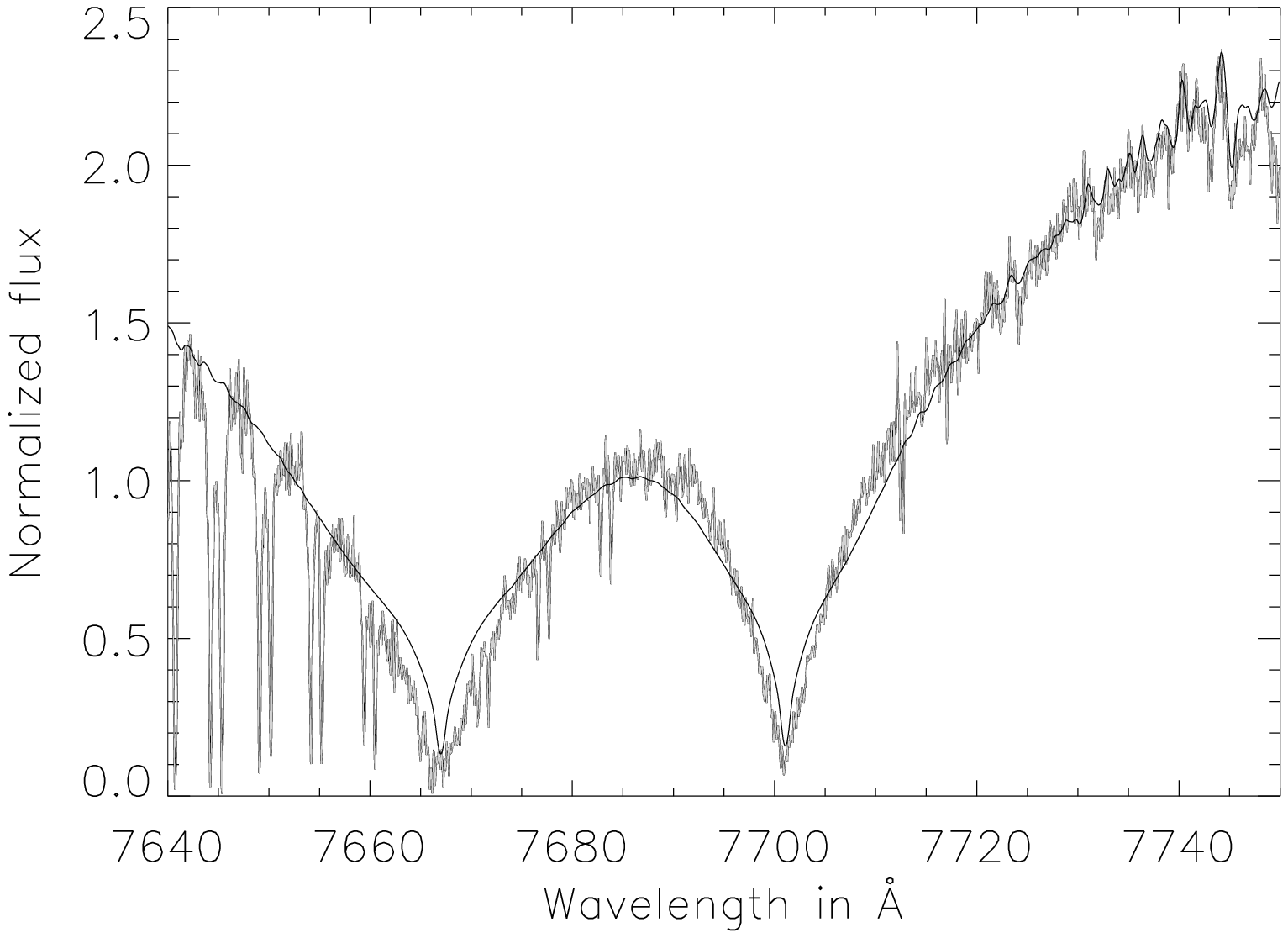}\\
\plotone
{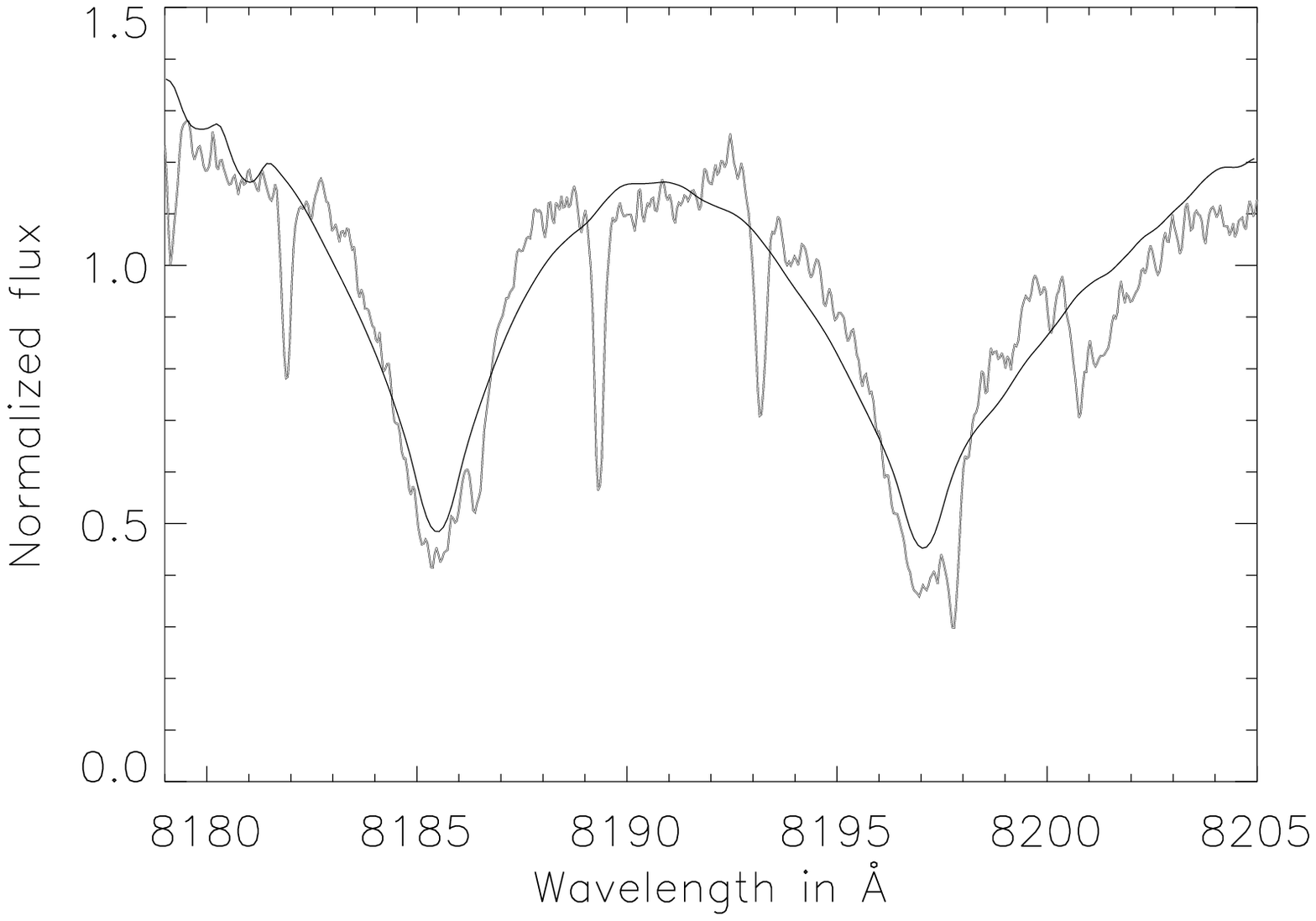}
\caption{\label{hi0746fit}
Fits (dark line) to 2M0746+200 (grey line).
See Tab. \ref{highresfittab} for parameters.
Telluric features have not been removed.
}
\end{figure}

\begin{figure}
\epsscale{0.49}
\plotone
{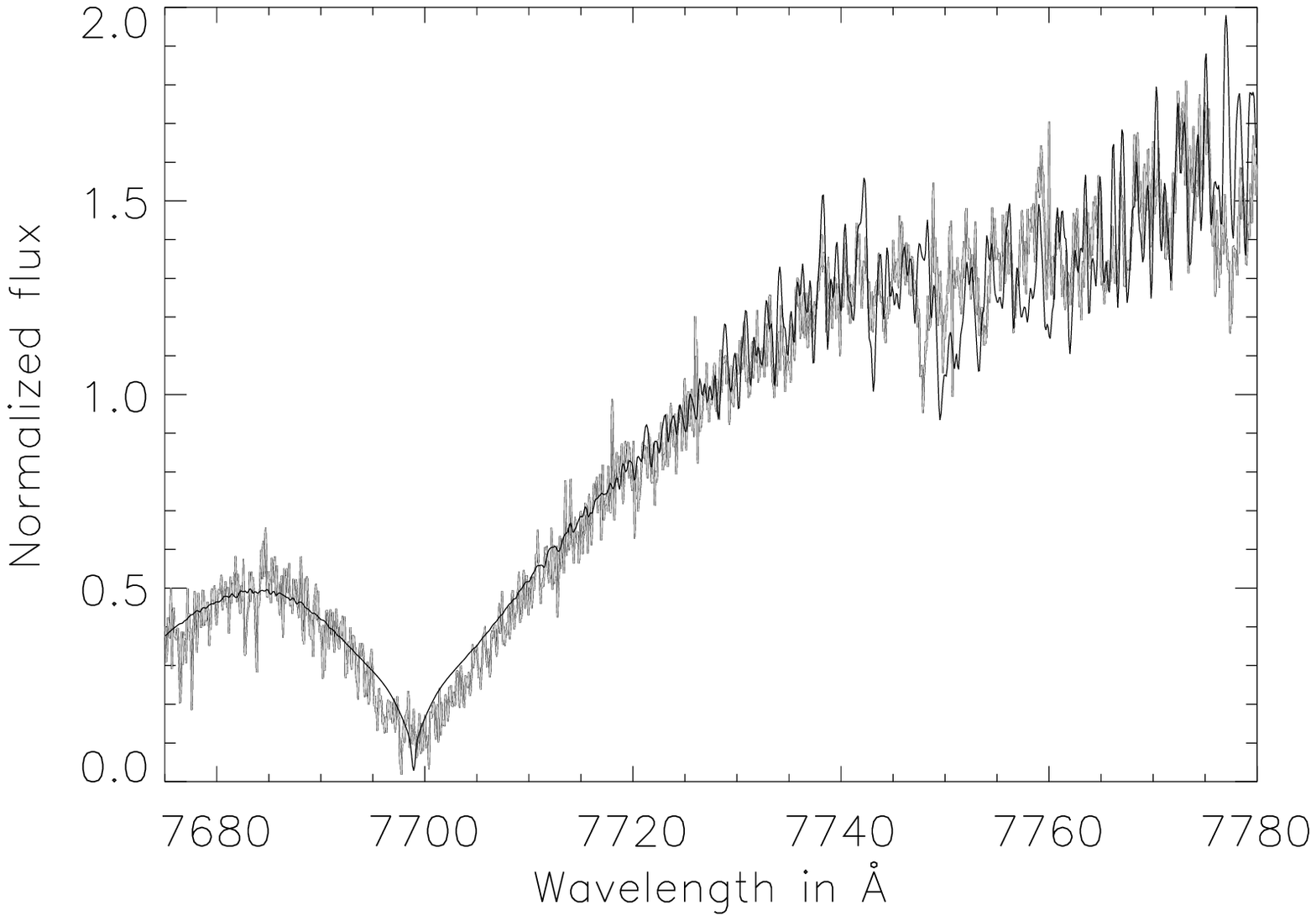}
\plotone
{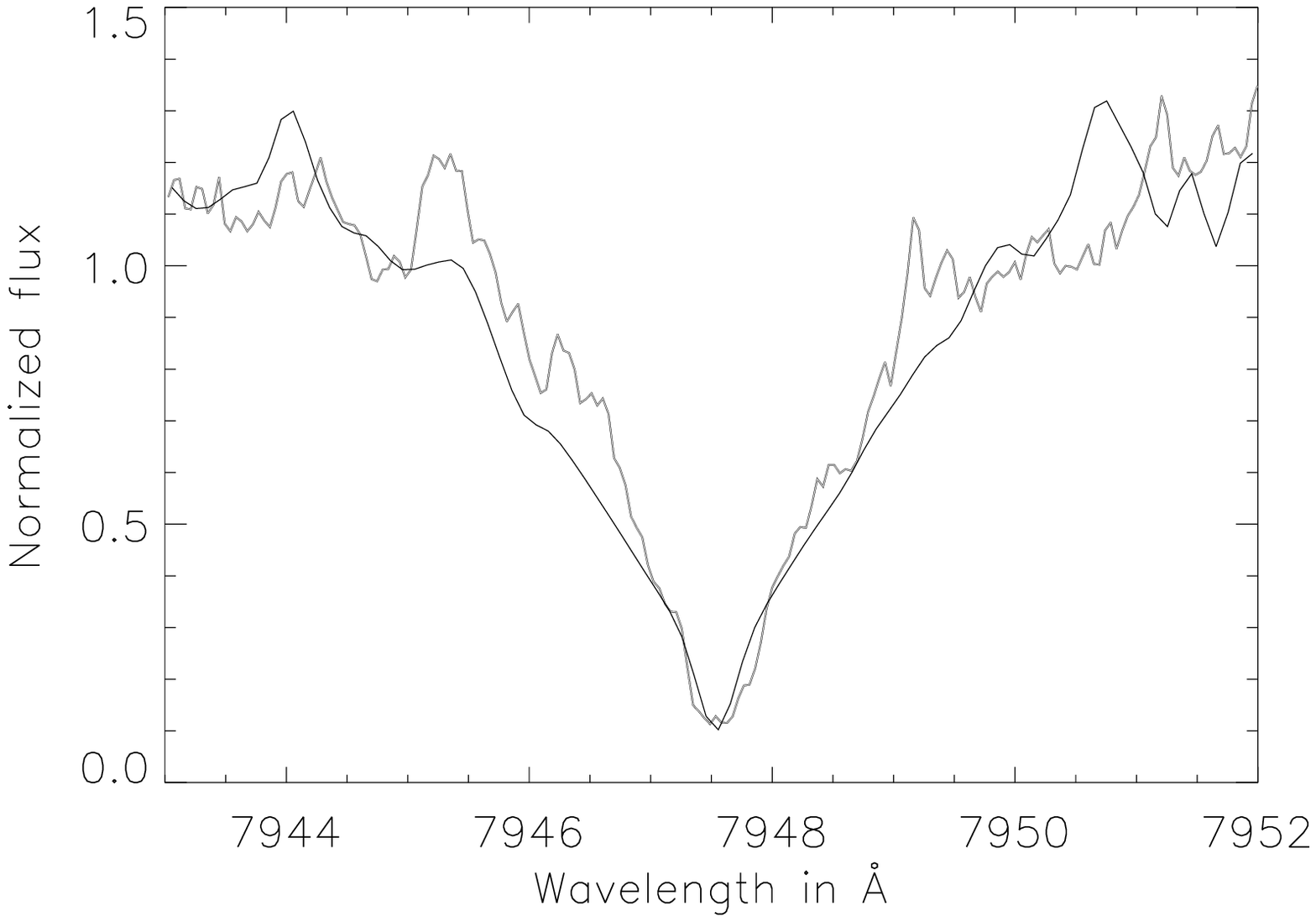}\\
\plotone
{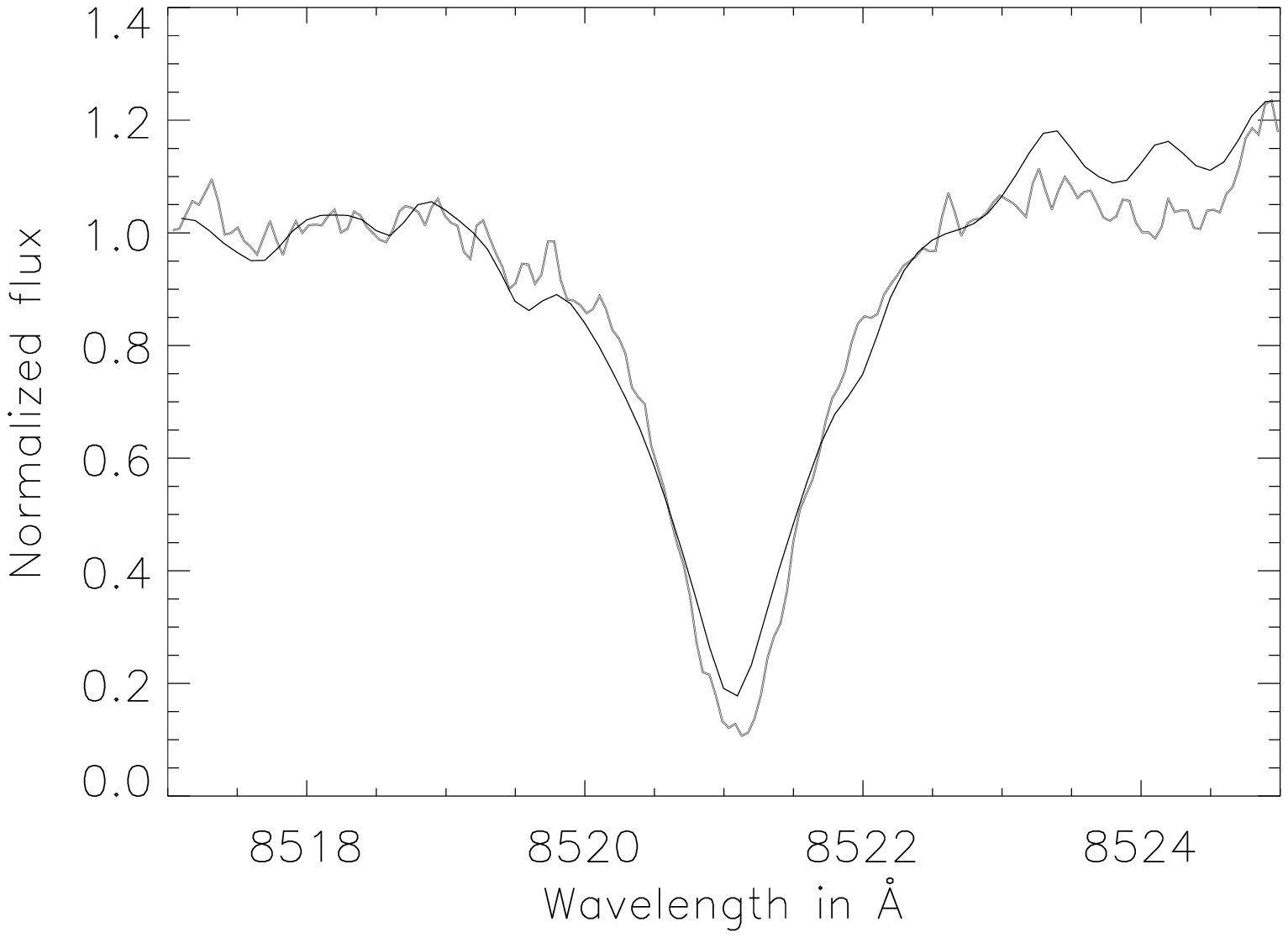}
\caption{\label{hi1439fit}
Fits (dark line) to 2M1439+1929 (grey line).
See Tab. \ref{highresfittab} for parameters.
Telluric features have not been removed.
}
\end{figure}

\begin{figure}
\epsscale{0.49}
\plotone
{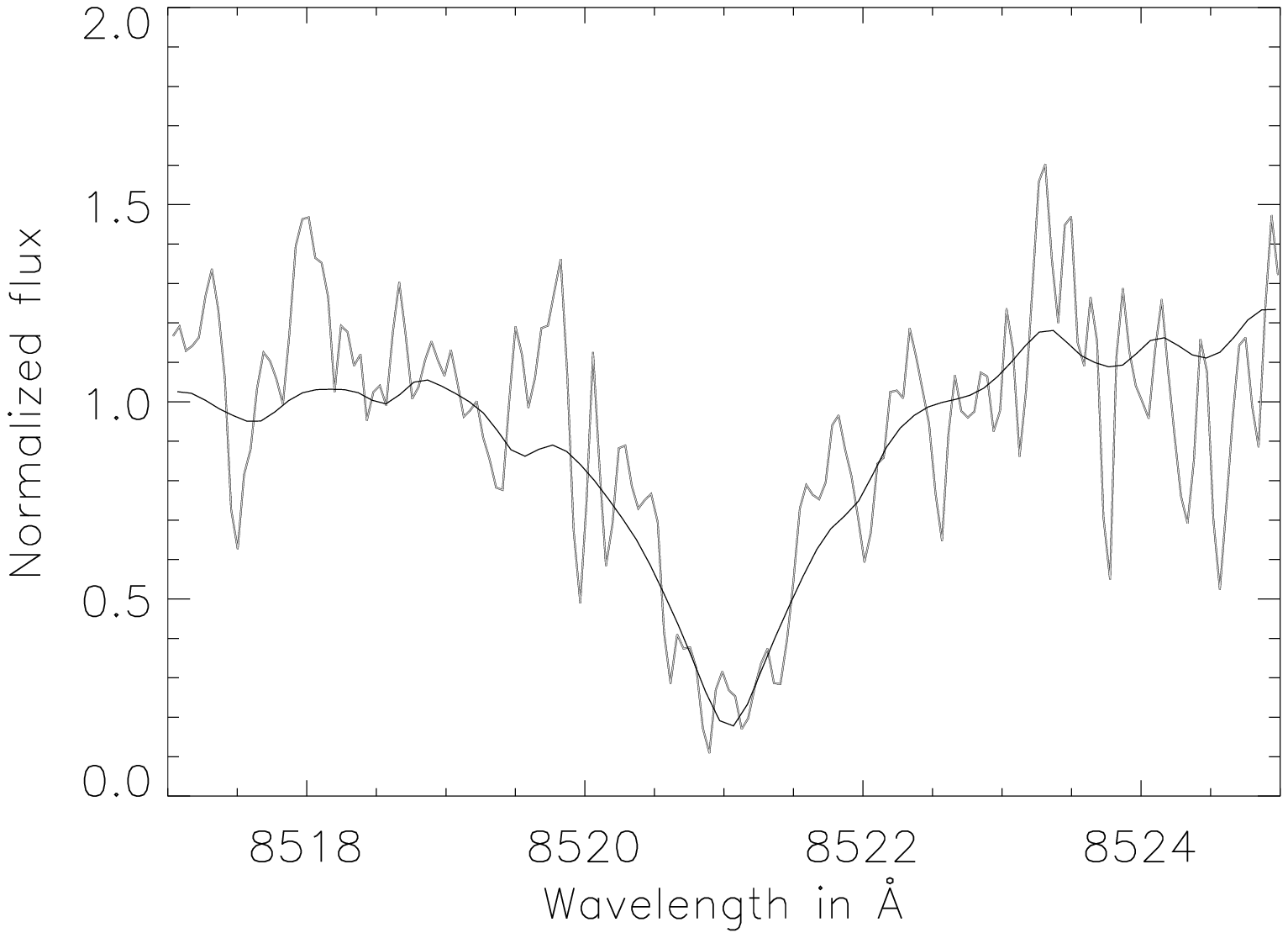}
\caption{\label{hi1726fit}
Fits (dark line) to 2M1726+1538 (grey line).
See Tab. \ref{highresfittab} for parameters.
Telluric features have not been removed.
}
\end{figure}

\begin{figure}
\epsscale{0.49}
\plotone
{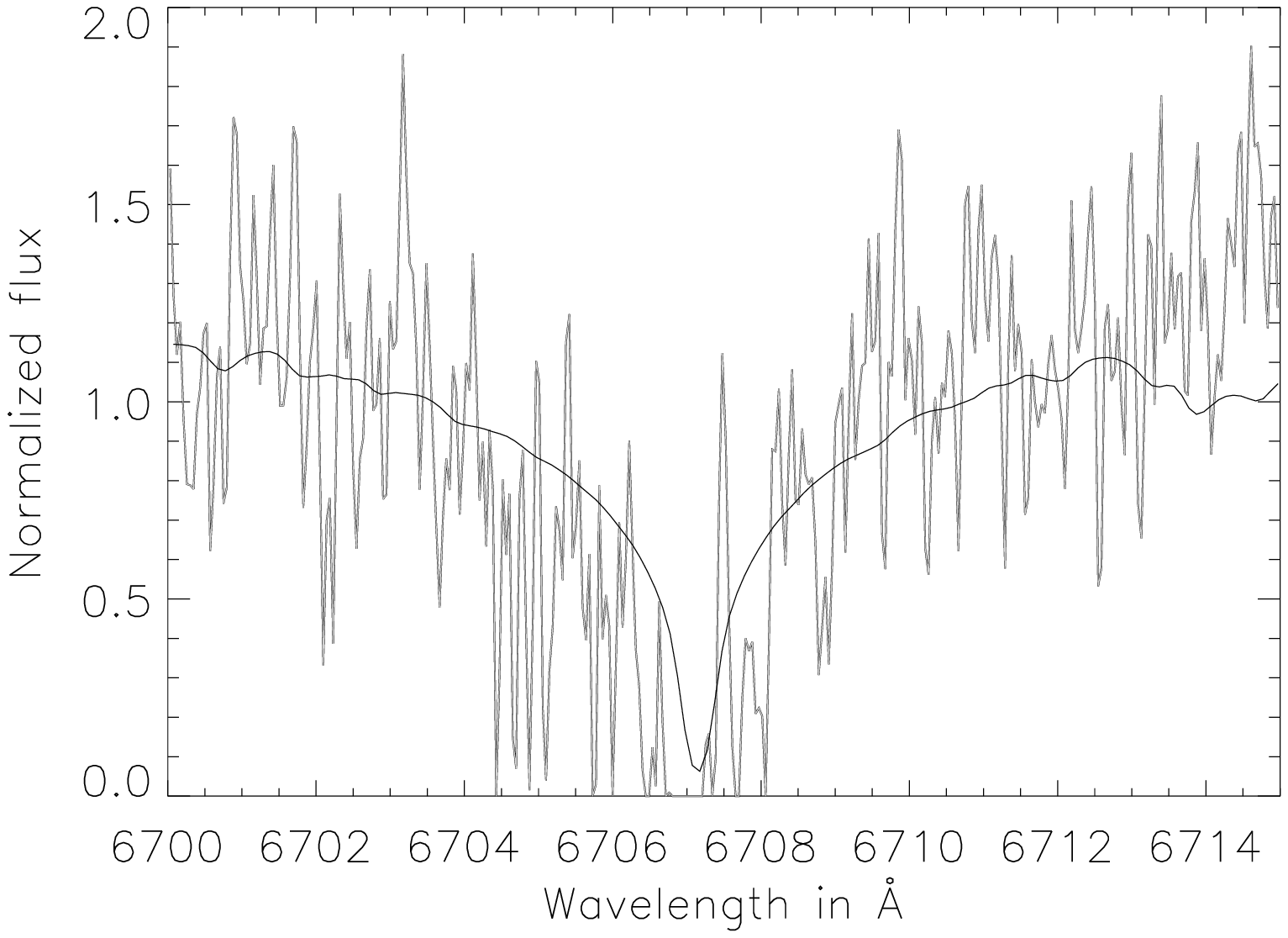}
\plotone
{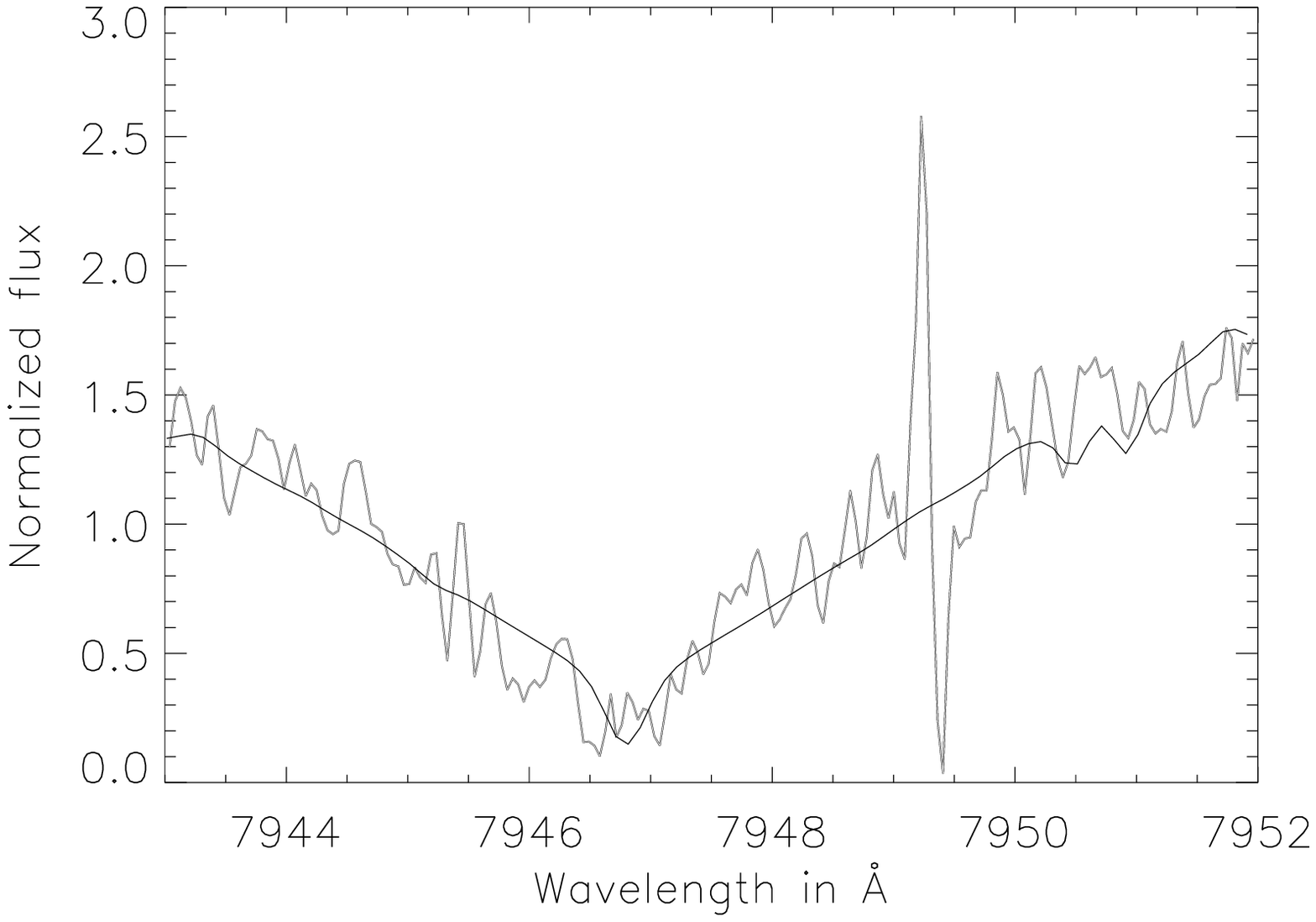}\\
\plotone
{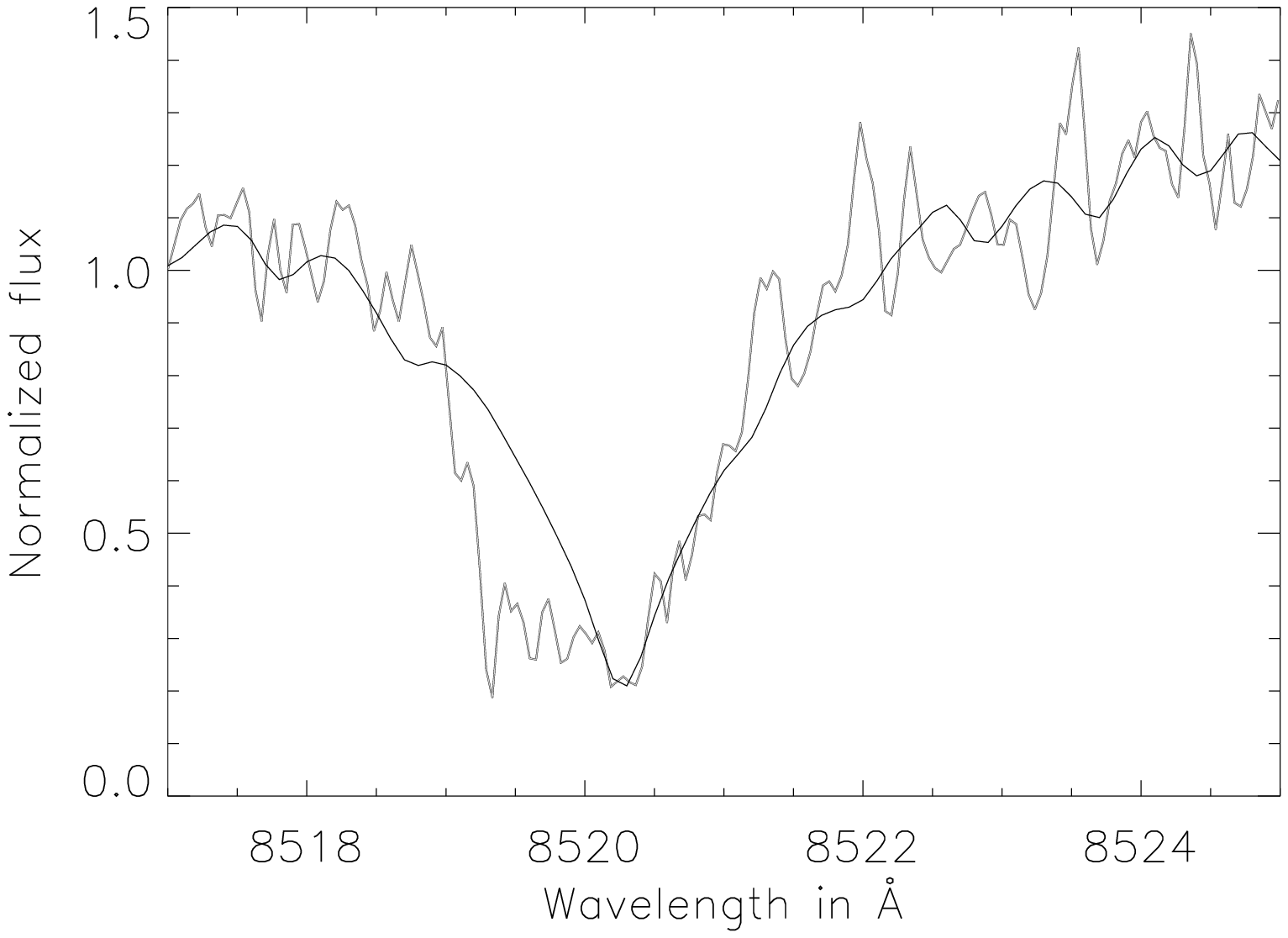}
\caption{\label{hi1146fit}
Fits (dark line) to 2M1146+2230 (grey line).
See Tab. \ref{highresfittab} for parameters.
Telluric features have not been removed.
}
\end{figure}

\begin{figure}
\epsscale{0.49}
\plotone
{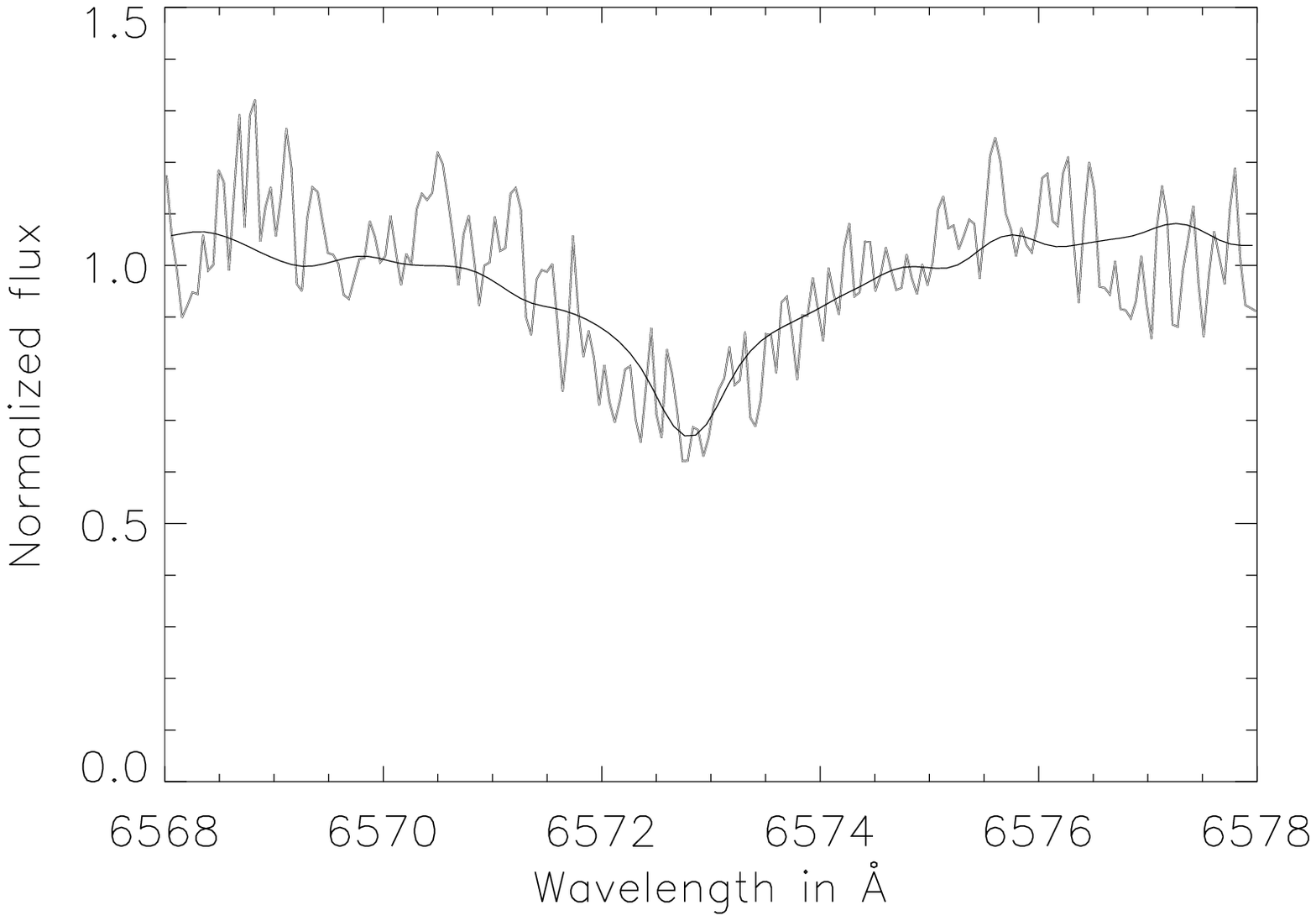}
\plotone
{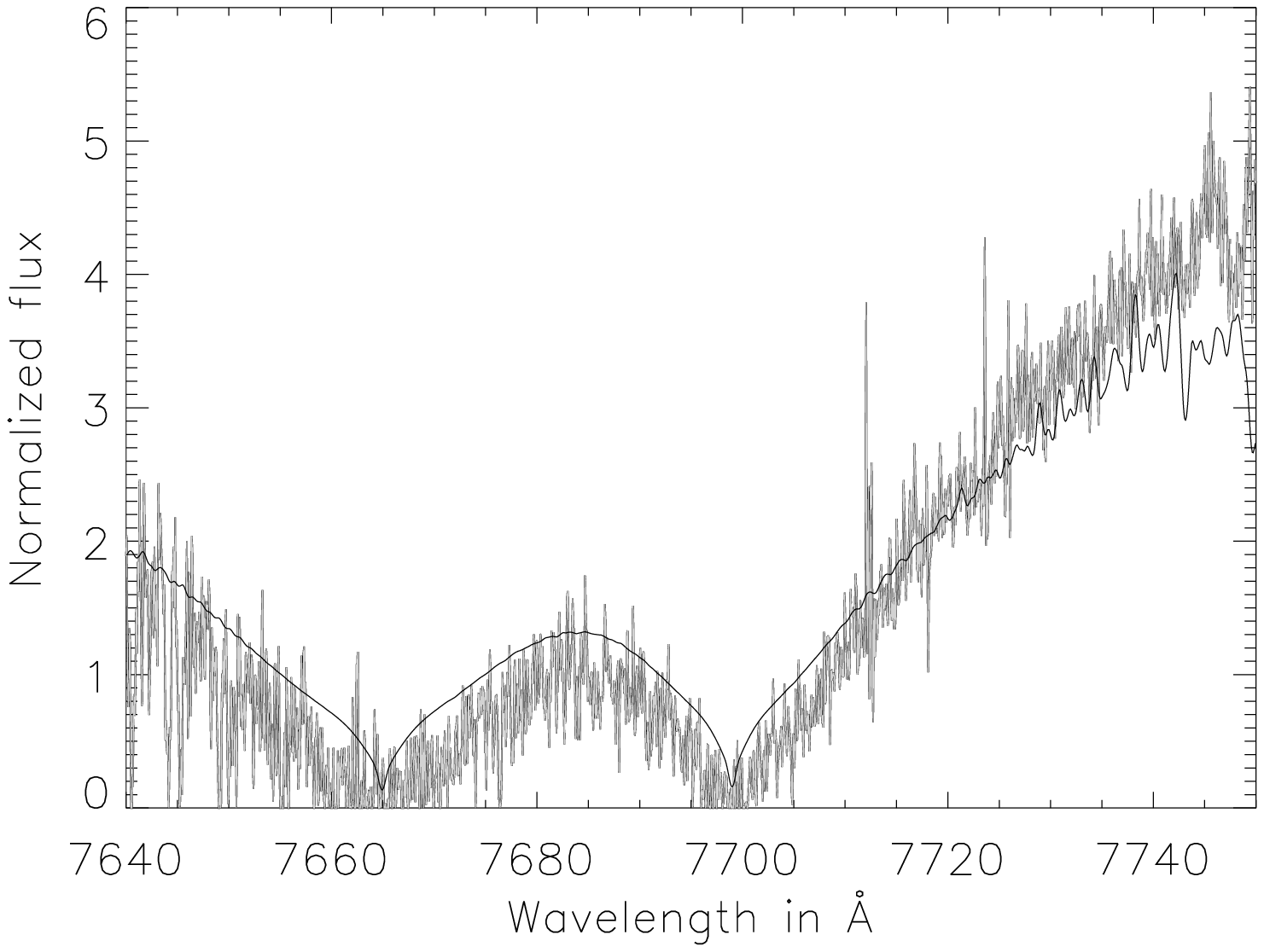}\\
\plotone
{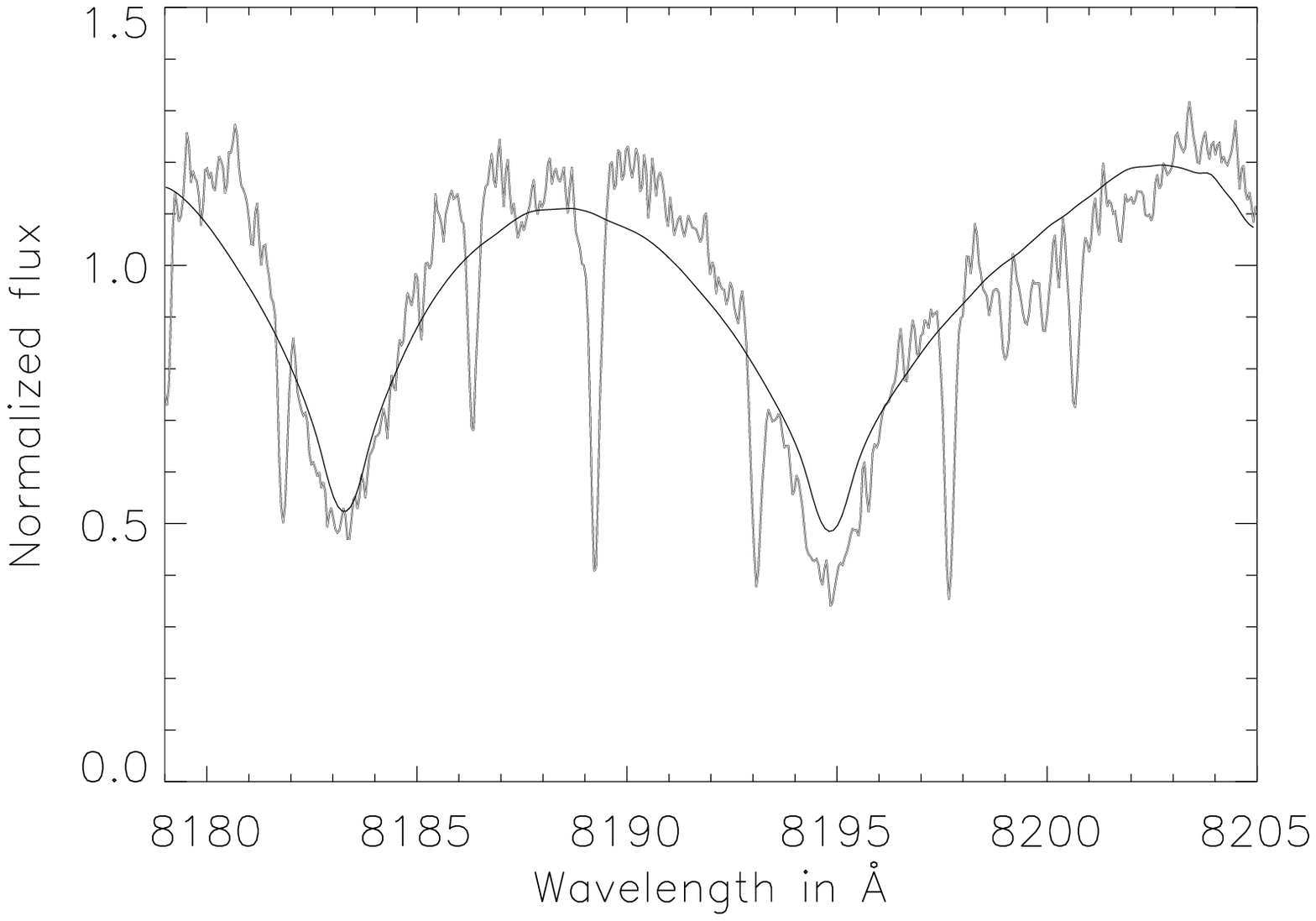}
\plotone
{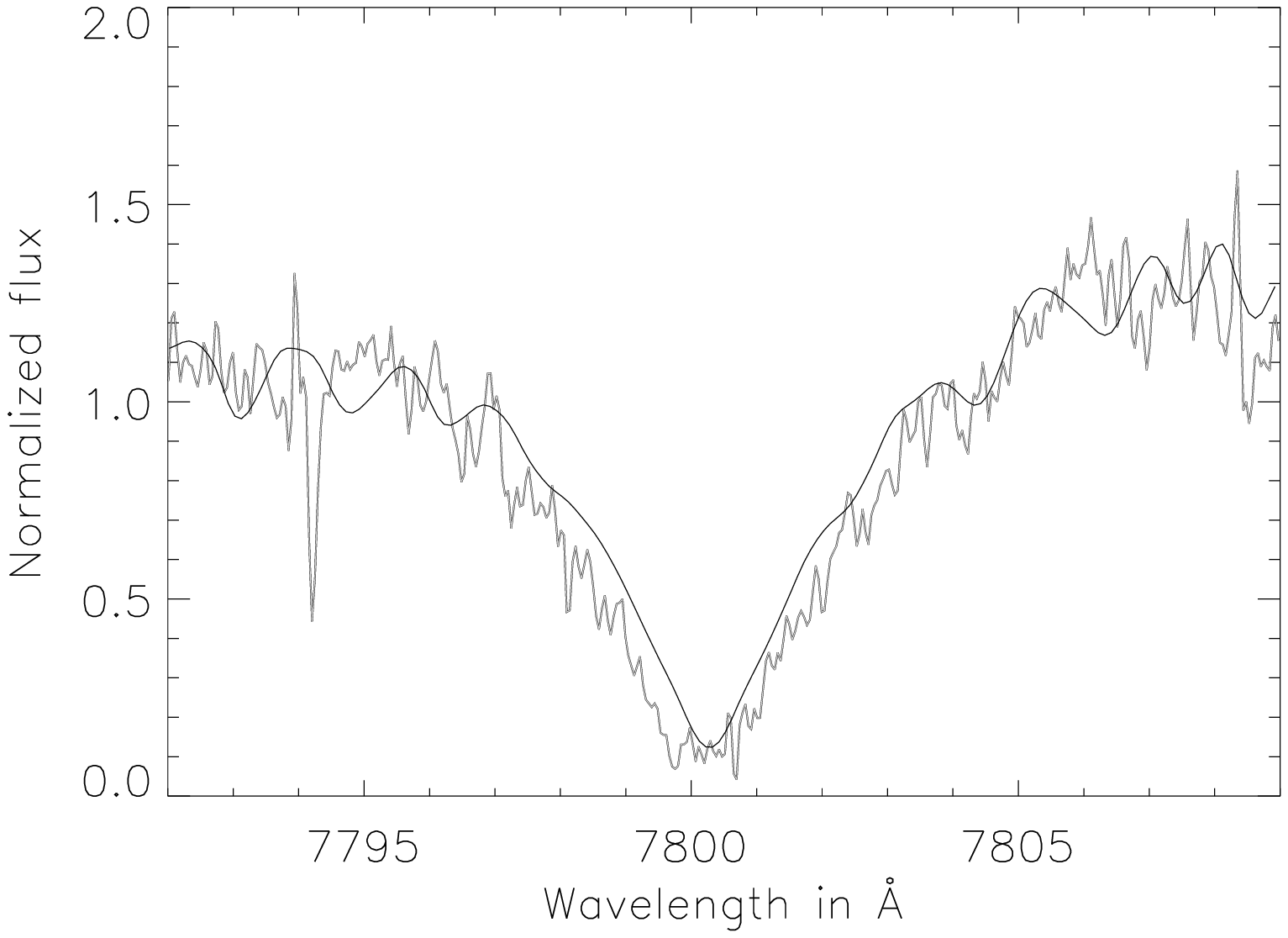}\\
\plotone
{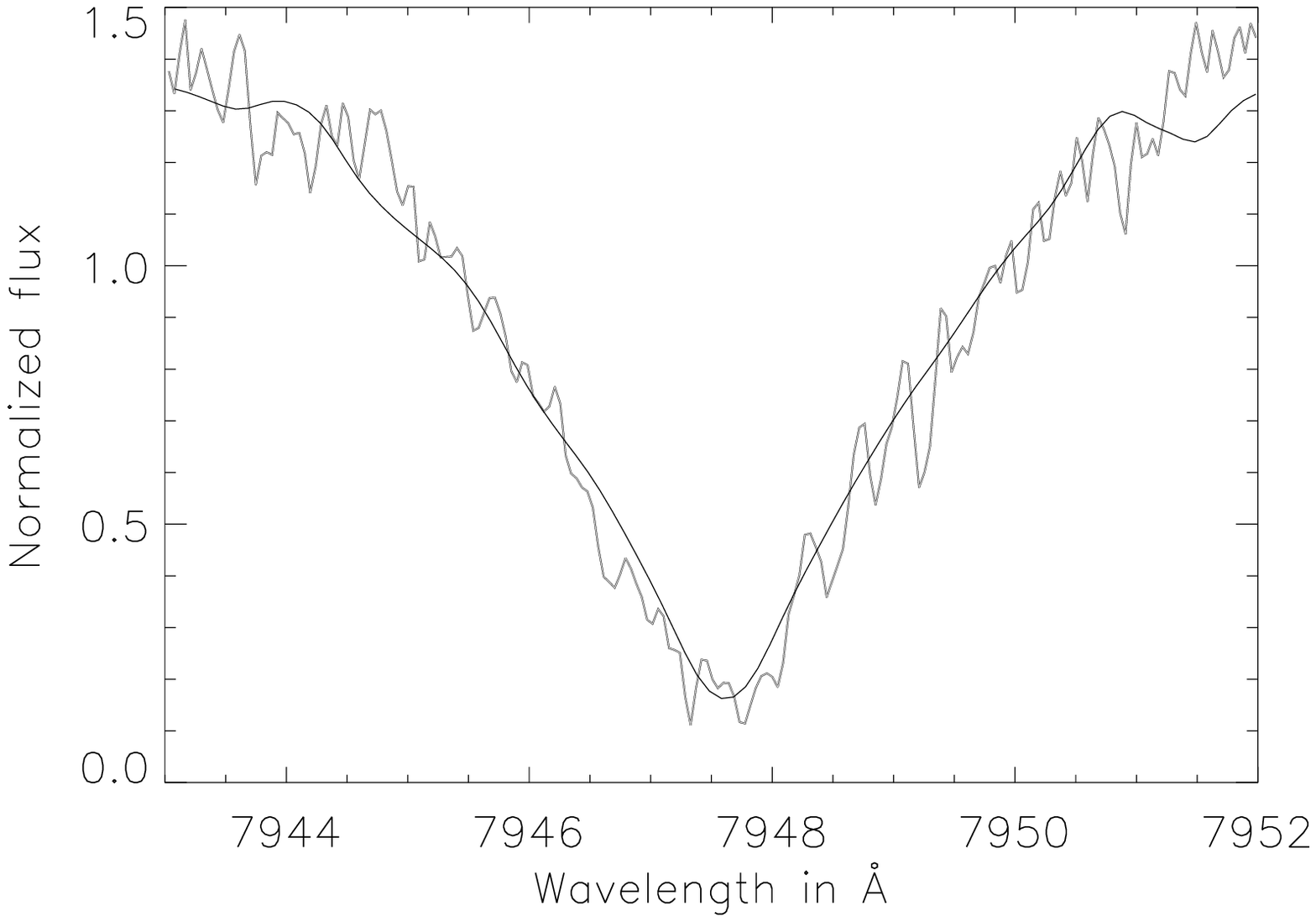}
\plotone
{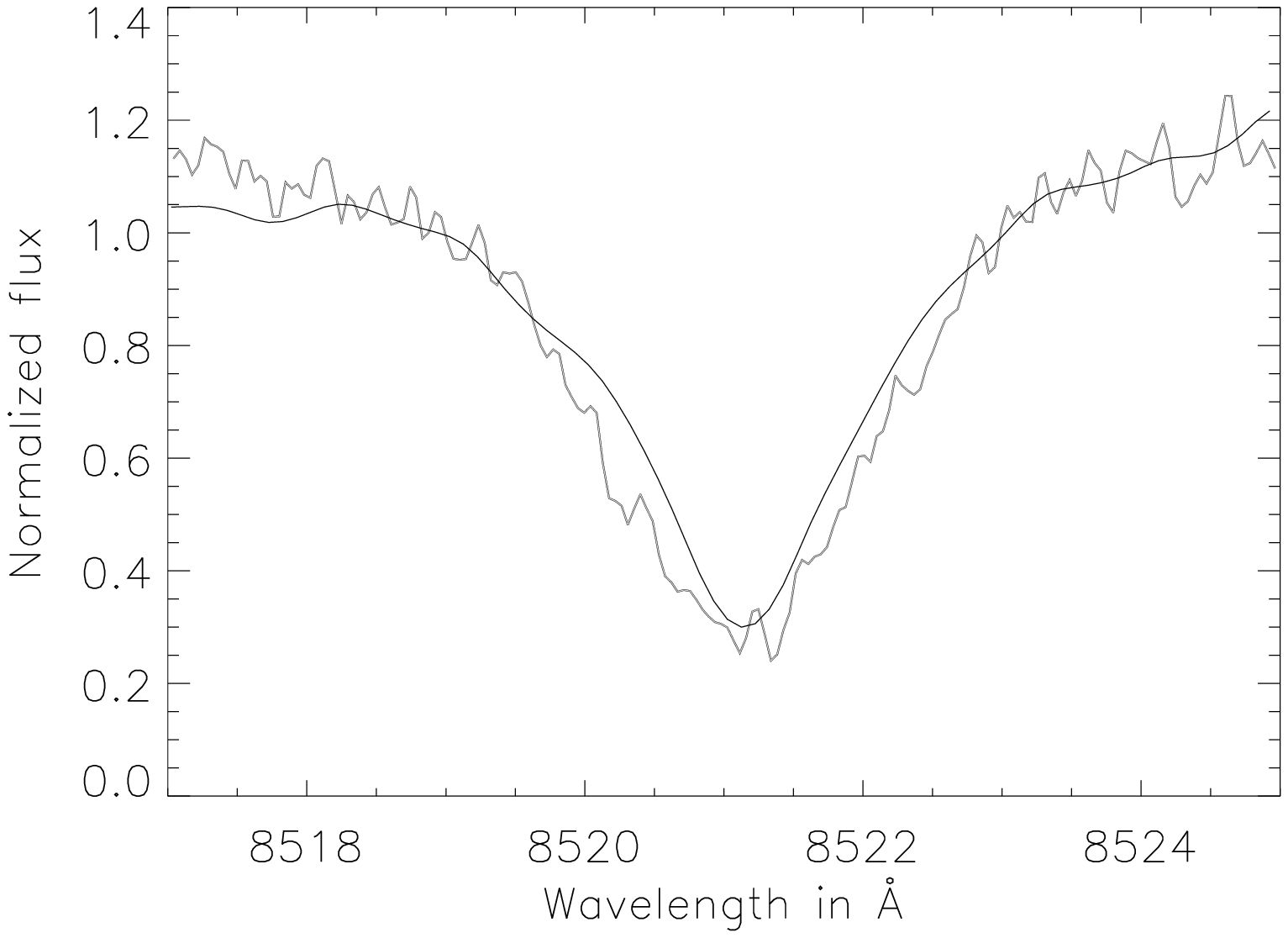}
\caption{\label{hi0036fit}
Fits (dark line) to 2M0036+1821 (grey line).
See Tab. \ref{highresfittab} for parameters.
Telluric features have not been removed.
}
\end{figure}

\begin{figure}
\epsscale{1.0}
\plotone
{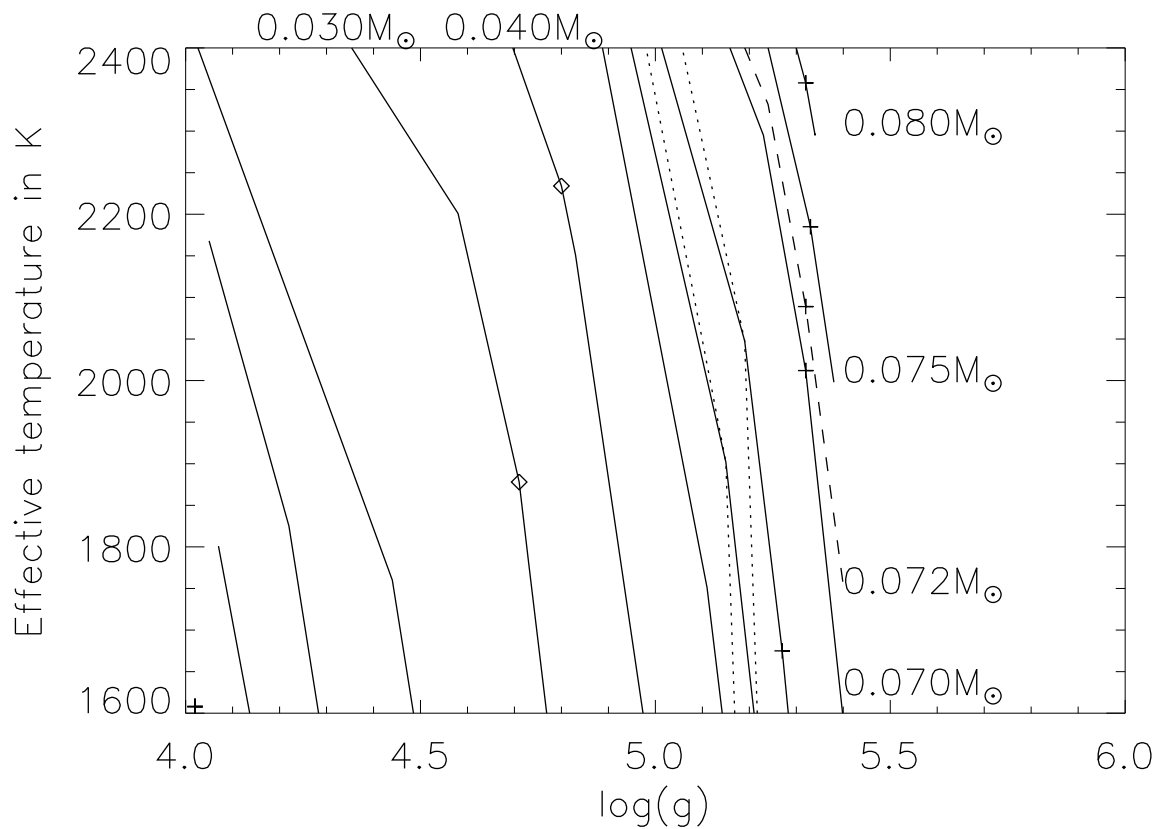}
\caption{\label{teff_logg}
\teff\ versus \logg\ from the data given by \citet{chabrier2000}.
Each line represents the track for a certain mass.
Some masses are indicated on the right and on the top.
The dashed line is the minimum stellar mass.
The dotted lines represent 95\% Li abundance remaining (left line) and
50\% Li abundance remaining (right line).
The diamonds are the points where the objects are 100~Myrs old, the crosses
where they are 1~Gyr old.
}
\end{figure}

\clearpage

\begin{deluxetable}{llrlrl}
\tablecaption{Parameters from the low resolution fits for the objects in the sample. \label{lowresfittab}}
\tablewidth{0pt}
\tablehead{
\colhead{} & \colhead{} & \multicolumn{2}{c}{AMES-Dusty} & \multicolumn{2}{c}{AMES-Cond}\\
\colhead{Name} & \colhead{Spectral Type} & \colhead{\teff}   & \colhead{\logg} & \colhead{\teff}   & \colhead{\logg}  }
\startdata
2M0149+2956   &  M9.5     & 2100  &  6.0  &  2100  &  6.0 \\
2M2234+2359   &  M9.5     & 2000  &  6.0  &  2000  &  6.0 \\
2M0345+2540   &  L0       & 1900  &  6.0  &  1800  &  5.5 \\
2M0147+3453   &  L0.5     & 1900  &  6.0  &  1800  &  5.5 \\
2M0746+200    &  L0.5     & 2000  &  6.0  &  2000  &  6.0 \\
2M1439+1929   &  L1       & 1900  &  6.0  &  1800  &  5.5 \\
2M1726+1538   &  L2       & 1900  &  6.0  &  1900  &  6.0 \\
2M1146+2230   &  L3       & 1800  &  5.0  &  1800  &  5.5 \\
2M0036+1821   &  L3.5     & 1800  &  5.5  &  1800  &  5.5 \\
\enddata

\end{deluxetable}

\begin{deluxetable}{llrlrlrlrlrlrlr}
\tabletypesize{\scriptsize}
\rotate
\tablecaption{Parameters from the high resolution fits for the objects in the sample. \label{highresfittab}}
\tablewidth{0pt}
\tablehead{
\colhead{} & \colhead{} &
\multicolumn{2}{c}{\ion{K}{1} 7685\ang} &
\multicolumn{2}{c}{\ion{Rb}{1} 7800\ang} &
\multicolumn{2}{c}{\ion{Rb}{1} 7948\ang} &
\multicolumn{2}{c}{\ion{Na}{1} 8190\ang} &
\multicolumn{2}{c}{\ion{Cs}{1} 8521\ang} &
\multicolumn{2}{c}{\ion{Ca}{1} 5673\ang} & \colhead{} \\
\colhead{Name} & \colhead{Spectral Type} & \colhead{\teff}   & \colhead{\logg} & \colhead{\teff}   & \colhead{\logg}  & \colhead{\teff}   & \colhead{\logg} & \colhead{\teff}   & \colhead{\logg} & \colhead{\teff}   & \colhead{\logg} & \colhead{\teff}   & \colhead{\logg} & v$_{\rm rot}$ }
\startdata
2M0149+2956   &  M9.5    &  2000  &  6.0  &  2000  &  6.0  &  2000  &  6.0  &  2000  &  6.0  &  2000  &  5.0  &        &       & 10 \\
2M2234+2359   &  M9.5    &  1900  &  5.0  &        &       &        &       &        &       &        &       &        &       & 15 \\
2M0345+2540   &  L0      &  2100  &  6.0  &  1900  &  5.0  &  1800  &  6.0  &  1800  &  5.5  &  1900  &  5.5  &        &       & 25 \\
2M0147+3453   &  L0.5    &  2000  &  5.0  &        &       &  1900  &  6.0  &        &       &  2000  &  5.0  &        &       & 10--20\\
2M0746+200    &  L0.5    &  2000  &  5.0  &        &       &        &       &  1900  &  5.0  &        &       &  1900  &  5.5  & 20  \\
2M1439+1929   &  L1      &  2100  &  5.0  &        &       &  1900  &  4.5  &        &       &  2100  &  5.5  &        &       &  \\
2M1726+1538   &  L2      &        &       &        &       &        &       &        &       &  2100  &  5.5  &        &       & \\
2M1146+2230   &  L3      &  &  $\ge$ 5.0  &        &       &  2100  &  6.0  &        &       &  2000  &  5.5  &  2000\tablenotemark{a}  &  5.5\tablenotemark{a} \\
2M0036+1821   &  L3.5    &  2100  &  5.0  &  2100  &  5.0  &  2000  &  5.0  &  1900  &  5.5  &  2100  &  6.0  &  1800  &  5.0  & 15 \\
\enddata

\tablenotetext{a}{These are from the \ion{Li}{1} 6708\ang\ line.}
\end{deluxetable}

\end{document}